\documentclass[]{jfm}

\usepackage{bm} 
\usepackage{amssymb}
\usepackage{natbib}
\usepackage{graphicx}
\usepackage{color}
\usepackage{citesort}

\usepackage{amsmath}

\newcommand{\G}{{\cal G}}
\newcommand{\x}{{\bm x}}

\newcommand{\be}{\begin{equation}}
\newcommand{\ee}{\end{equation}}

\begin{document}

\title[Wetting failure and contact line dynamics in a Couette flow]{Wetting failure and contact line dynamics in a Couette flow}

\author[M. Sbragaglia, K. Sugiyama and L. Biferale]
{M.\ns S\ls B\ls R\ls A\ls G\ls A\ls G\ls L\ls I\ls A$^1$, K.\ns S\ls U\ls G\ls I\ls Y\ls A\ls M\ls A$^2$, and L.\ns B\ls I\ls F\ls E\ls R\ls A\ls L\ls E$^{1}$}

\affiliation{$^1$Department of Physics and INFN, University of Tor Vergata, \\ Via della Ricerca Scientifica 1, 00133 Rome, Italy \\
$^2$ Fluid Engineering Laboratory, The University of Tokyo,\\ Department of Mechanical Engineering, 7-3-1 Hongo Bunkyo-Ku, Tokyo 113-8656, Japan}



\date{\today}
\maketitle

\begin{abstract}
Liquid-liquid wetting failure is  investigated in a two-dimensional Couette system with two immiscible fluids of arbitrary  viscosity.  The problem is solved exactly using a  sharp interface treatment of hydrodynamics (lubrication theory) as a function of the capillary number, viscous ratio and separation of  scale, i.e. slip length  versus macroscopic scale of the system.  The existence of critical velocities, above which no stationary solutions are found, is analyzed in detail in terms of the relevant parameters of the system. Comparisons with existing analysis for other geometries  are also carried out. A numerical  method of analysis  is also presented, based on diffuse interface models obtained from multiphase extensions of the lattice Boltzmann equation (LBE). Sharp interface and diffuse interface models are quantitatively compared face to face indicating the correct limit of applicability of the diffuse interface models.
\end{abstract}

\section{Introduction}

Despite many years of research, the physics of moving contact lines (\cite{DeGennes85,Oron97,Blake06}) is still not completely understood. This lack of understanding stems from different factors. Dynamical wetting operates on scales extending from the macroscopic to the molecular ones. In between these scales, with the usually small Reynolds numbers achieved in such motion, the strong viscous forces are balanced with surface tension effects (\cite{DeGennes85,Voinov76,Cox86}). A dimensionless measure of this balance is provided by the capillary number $Ca=\mu U/\sigma$, comparing the viscous term at the contact line, $\eta U$, with the surface tension, $\sigma$, where $\eta$ denotes the liquid viscosity and $U$ the contact line velocity. Liquid motion at finite capillary numbers induces changes in the shape of the interface, and the resulting macroscopic dynamic angle, $\theta_{M}(Ca)$, is different from its static equilibrium counterpart. Within this context, the main issue in contact line research is therefore to relate the macroscopic angle to the inner physics, with particular emphasis on the mechanisms releasing the small scale singularities (\cite{Cox86,DeGennes86,Voinov76}). In fact, it is  well known that the viscous stress diverges at the contact line if some  physical mechanism is not introduced to remove the singularity (\cite{Dussan79,DeGennes86,Pismen00}). To overcome this problem, numerous  proposals came out (\cite{HuhScriven71,Voinov76,HockingRivers82,DeGennes85,Cox86,Dussan91}), all of them conducting to  the introduction of a small length scale parameter $\ell_s$ used to release the viscous singularity. An example is provided by the slip length at the boundaries, where this small scale parameter is related to the presence of a finite slip (\cite{Cox86,HuhScriven71}). Anyhow, also other mechanisms such as intermolecular forces (\cite{DeGennes86,Pismen00,Jacqmin00,Spelt07}) in the immediate vicinity of the contact line can be considered.

Once the small scale singularity is removed, the macroscopic angle emerges as a function of the capillary number $\theta_{M}(Ca)$: although the explicit forms may clearly depend on the specific model used to release the singularity, the common feature is that the small capillary number limit scales linearly with the contact line velocity, $\theta_{M}(Ca) \sim Ca$ (\cite{Voinov76,Cox86}).  Differences are expected to emerge close to the wetting transition: when the liquid advances, a critical speed exists above which a stationary contact line cannot be sustained any longer, and liquid deposition may occur on the solid (\cite{Eggers04,Jacqmin04,Snoeijer07,Blake79,Quere91,Podgorski01,DeGennes86,Sedev91,Simpkins03}). The understanding of this transition is crucial. In fact, the breaking of stationarity can be interpreted as a lost of universality: the hydrodynamical regime does not support anymore a time-independent solution, indicating some singular behavior in the matching between inner and outer regions. 

In his review, \cite{Kistler93} supports the assumption that wetting failure occurs when the dynamic angles reach zero degrees \footnote{Throughout the paper notations are consistent with a macroscopic angle that decreases as the capillary number increases (see also the geometry in figure \ref{fig:1} with $\theta_{M}$ smaller than $\theta_{m}$).}, whereas the contact line is observed to become V-shaped in the vicinity of the instability (\cite{Blake79,Ghannum93}).

Evidence for the existence of such critical points of entrainment has also been provided by some recent theoretical works. The problem has been tackled using a full hydrodynamic calculation incorporating viscous effects on all scales (\cite{Eggers04,Eggers05,Hocking01}). In particular, within the framework of the 'Landau-Levich' problem (\cite{Landau42,Derjaguin43}), i.e. a plate plunging into or being withdrawn from a liquid bath, it has been shown  (\cite{Eggers04,Eggers05})  that stationary solutions cease to exist above a critical capillary number $Ca_{cr}$.  The value of the emerging  critical capillary number is  quantified exactly: although we find universal features in terms of the microscopic angle ($\theta_m$) of the liquid at the wall ($Ca_{cr} \sim \theta_{m}^3$), a non-universality is also present in the geometrical prefactor depending on the angle of inclination of the plate with respect to the liquid (\cite{Eggers04,Eggers05}).  

The whole picture is also enriched by recent experimental observations (\cite{Snoeijer06}) where it is shown that the formation of  solitary capillary waves can drastically change the value of the critical speeds of entrainment (pre-critical wetting transition). Along these lines, linear stability analysis (\cite{Golestanian01a,Golestanian01b,Snoeijer07}) for the  relaxation to external perturbations  has also revealed that dispersion relations behave differently away and close to the critical point. 

Wetting failure has also been investigated in a  recent paper by \cite{Jacqmin04}, where a liquid-liquid iso-viscous system in a Couette geometry was treated using various methods. This system, consisting of two walls moving with opposite velocities, has been treated in the parallel flow approximation, with Fourier series method and also with phase field models (\cite{Jacqmin00,Jacqmin04}). In these liquid-liquid systems, wetting failure has been found to set in well before the dynamic angle reaches zero degrees.

In this paper, we further explore and elaborate on these issues by following a double-sided strategy.  First, we  extend the  analysis of \cite{Jacqmin04} to arbitrary  viscosity ratios, $\chi = \mu_g/\mu_l$, where $\mu_{g}$ and $\mu_{l}$ are a gas and liquid viscosity respectively. We then quantify the breaking of a stationary regime  at a critical capillary number, $Ca_{cr}(\chi,\lambda,\theta_m)$, which depends on the  viscosity ratio, $\chi$, the microscopic wettability, $\theta_m$, and the ratio between inner and outer scales, $\lambda=\ell_{s}/H$, where $\ell_s$ is the slip length  associated to a finite contact line slip and  $H$ is  the distance between the two plates in the Couette geometry (see section \ref{sec:COUETTE}). The theoretical approach used is based on the lubrication approximation, i.e. a sharp interface treatment of hydrodynamics dealing with weakly bended interfaces (\cite{Oron97}). In order to investigate macroscale geometry effects we also compare our results to those of a similar analysis applied to the Landau-Levich problem (\cite{Eggers04,Eggers05}).

In the second part of the paper we will follow a computational pathway. We will study the same Couette flow with  multiphase extensions of the lattice Boltzmann equation (LBE) (\cite{Gladrow,Saurobook}). In these diffuse interface methods  effective slip is induced by the finite width of the interface (\cite{Seppecher96,Spelt07}) which is explicitly considered in the separation of two bulk phases. On one side, the LBE will serve as a benchmark of theory in those cases when the analytical approach is questionable (strongly bended interfaces  and/or finite contact angles).  On the other side, we will use the sharp interface  calculation to study the effects of finite width of the interface on the system. In fact, comparisons between a sharp interface theory (with slip-released singularity) and LBE (with diffuse-released singularity) is a way to probe the universality of the dynamics far from the contact line, i.e. that it does not depend on the microscopic details used to release the singularity.

Dynamical benchmarks of LBE for multi-phase flows are also important, because of the difficulties to control corrections to the hydrodynamical limit in presence of strong density gradients (\cite{Sbragaglia07,Shan06,Lee06,Wagner03,Cristea03,Yuan06}). We will show that LBE supports asymptotically the correct hydrodynamical behavior, but it may be necessary to look on macroscopic scales much larger than the interface width. The parameter controlling deviations from sharp interface hydrodynamics is therefore going to be proportional to the ratio between the inner length scale (i.e. interface width) and the outer scale of the system.

\section{Lubrication Approach in the Couette flow}\label{sec:COUETTE}

\begin{figure}
\begin{center}
\includegraphics[scale=0.7]{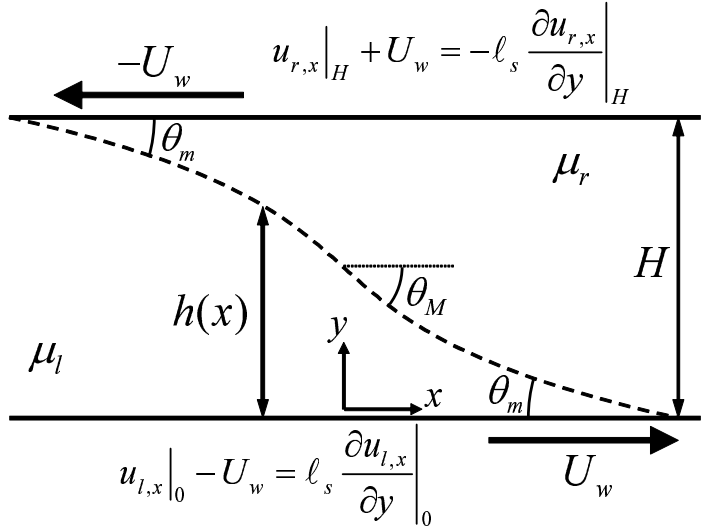}
\caption{The geometry used to analyze the stability problem. Two parallel
 walls are moved with opposite velocities $\pm U_{w}$. The two fluids under consideration are distinguished in left (l, more viscous) and right (r, less viscous) and the viscosity ratio will be denoted by $\chi=\mu_{r}/\mu_{l}$.
 The angles that the  fluids are forming with the walls are
 complementary (opposite wetting properties). The microscopic angle
 $\theta_{m}$ is defined as the angle that the left fluid is forming
 with the lower wall and the angle $\theta_{M}$ is taken as the angle in the center of the channel, with the convention that for finite capillary number $\theta_{M}$ is smaller that $\theta_{m}$. The interface is determined by a function $h(x)$ of the horizontal coordinate. Streamwise and vertical velocity fields for the left (right) fluid will be denoted with $u_{l,x}$ and $u_{l,y}$ ($u_{r,x}$
 and $u_{r,y}$).  }
\label{fig:1}
\end{center}
\end{figure}

The geometry we study consists of two parallel walls moving with opposite velocities $\pm U_{w}$. The two walls have opposite wettability so that when the fluid is motionless the interface is a straight line angling from wall to wall (see figure \ref{fig:1}). For each fixed horizontal \footnote{The origin of the coordinates is chosen so that the horizontal spreading of the interface is centered at $x=0$} location ($x$), the interface profile will be denoted with $h(x)$. Computations will be carried out under the assumption of a finite and fixed slip length ($\ell_{s}$), for simplicity the same on both walls. Other parameters are the capillary number, $Ca=U_{w}\mu_{l}/\sigma$, and the viscosity ratio, $\chi=\mu_{r}/\mu_{l}$. It is further assumed that the left fluid is the more viscous one so that $\mu_{l}>\mu_{r}$ and consequently $\chi \le 1$.  The computation at $\chi=1$ will serve as a benchmark test for our results when compared to the results given in the paper of \cite{Jacqmin04}. 

Using a lubrication approximation (\cite{Oron97,Jacqmin04,Eggers05,Eggers04,Snoeijer05}) we carry out a long wavelength expansion in the stationary equations of motion (i.e. Stokes equation and continuity equation). Therefore, the theory developed is  expected to be valid in the limit  of small tilting angles between the interface and the wall. In this way, for a fixed horizontal coordinate $x$, the complete set of equations  to be analyzed is 
\be\label{STOKES}
\mu_{l} \frac{\partial^{2} u_{l,x}}{\partial y^2}=\frac{\partial
  p_{l}}{\partial x} \hspace{.2in} \mu_{r} \frac{\partial^{2}
  u_{r,x}}{\partial y^2}=\frac{\partial p_{r}}{\partial x} \ee
\be\label{CCC}
\partial_{x} u_{l,x}+\partial_{y} u_{l,y}=0 \hspace{.2in} \partial_{x}
u_{r,x}+\partial_{y} u_{r,y}=0 
\ee 
where we label with $l$ ($r$) the left (right) fluid with viscosity $\mu_{l}$ ($\mu_{r}$) and pressure $p_{l}$ ($p_{r}$). These equations are the natural generalization of those analyzed in the paper of \cite{Jacqmin04}, where the author considered the case $\mu_l=\mu_r$. The solutions for the horizontal velocities in the left and right fluid  ($u_{l,x},u_{r,x}$) are rapidly evaluated as: 
\be
u_{l,x}=\frac{1}{\mu_{l}}\left(A_{l}+B_{l}y+\frac{1}{2}p_{l,x} y^2
\right) \hspace{.2in}
u_{r,x}=\frac{1}{\mu_{r}}\left(A_{r}+B_{r}y+\frac{1}{2}p_{r,x} y^2
\right) 
\ee 
where we have used $p_{l,x}=\frac{\partial p_{l}}{\partial x}$ and $p_{r,x}=\frac{\partial p_{r}}{\partial x}$. The vertical components $u_{l,y}$ and $u_{r,y}$ have to be evaluated from (\ref{CCC}) with the usual boundary condition of zero normal velocity at the wall. Obviously, $A_{l}$, $B_{l}$, $p_{l,x}$, $A_{r}$, $B_{r}$, $p_{r,x}$ have to be determined upon the imposition of ad-hoc boundary/matching  conditions that will determine the position of the interface $h(x)$. The relevant matching conditions are the continuity of the parallel velocity and viscous stress at the interface  
\be
\label{CONT} \left. u_{l,x}
\right|_{h} =\left. u_{r,x} \right|_{h} \hspace{.2in} \left. \mu_{l}
  \frac{\partial u_{l,x}}{\partial y} \right|_{h}=\mu_{r}
\left. \frac{\partial u_{r,x}}{\partial y} \right|_{h} 
\ee 
plus the lower and upper wall boundary conditions written as
\be\label{SLIP}
\left. u_{l,x} \right|_{0}-U_{w}=\ell_{s}\left. \frac{\partial
    u_{l,x}}{\partial y} \right|_{0} \hspace{.2in} \left. u_{r,x}
\right|_{H}+U_{w}=-\ell_{s}\left. \frac{\partial u_{r,x}}{\partial y}
\right|_{H} .  \ee 
Finally, we have the kinetic condition of no flux across the interface 
\be 
\hat{n}_{x} u_{l,x}+\hat{n}_{y}
u_{l,y}=\hat{n}_{x} u_{r,x}+\hat{n}_{y} u_{r,y}=0 
\ee
with $\hat{n}$ the normal at the interface in $h(x)$. In Appendix A  we show how to map the $6$ boundary conditions in a closed system  and solve the corresponding matrix problem as a function of the separation of scale, $\lambda=\ell_{s}/H$, the capillary  number, $Ca=U_{w}\mu_{l}/\gamma$, and  the viscosity ratio, $\chi=\mu_{r}/\mu_{l}$.\\
Once the pressure drop across the interface is known, one may derive using the Laplace law the equation for the local curvature, $\kappa$, of the interface
\be
\label{Xder} \sigma \frac{d \kappa}{d  x}=p_{r,x}-p_{l,x} 
\ee 
with $\sigma$ the surface tension at the interface. As already  noticed by \cite{Jacqmin04}, if we
define the interface arclength coordinate as $s$, we can write consistently with the lubrication approximation that 
\be
\label{Sder}
\sigma \frac{d \kappa}{d s }=p_{r,x}-p_{l,x} 
\ee 
where the derivative of the curvature is connected to the angular variation $\frac{d \tilde{\theta}}{d s}=\kappa$ (where
$\tilde{\theta}=\pi-\theta$).  Summarizing, the governing equation set is:
\be 
\label{eq:ode} \frac{d}{d s} (\kappa,\tilde{\theta},x,y)=(\sigma^{-1}(p_{r,x}-p_{l,x}),-\kappa,\cos
\tilde{\theta},-\sin \tilde{\theta}). 
\ee 
The determination of the interface becomes now a non-linear boundary value problem: we need to solve the ODE (\ref{eq:ode}) for a given capillary number, $Ca$, viscosity ratio, $\chi$, and separation of scale,  $\lambda=\ell_s/H$, with the boundary conditions for the microscopic angle at the wall equal to $\theta_m$ \footnote{for the present study we assume that the microscopic wall wettability is not dependent on the velocity}.
We have therefore a four-parameter problem.  To solve numerically the previous set of non-linear ODE, we adopt a second order Runge-Kutta method with a non uniform grid with increasing resolution near the wall. Notice that in presence of two different viscosities ($\mu_r \neq \mu_l$) the interface is not going to be symmetric with respect to the center of the channel. For a  given value of viscous ratio $\chi$ and capillary number $Ca$, we look for solutions of (\ref{eq:ode}) by fixing the angle in the center of the cell, $\theta_M$, and choosing the curvature in that location so as to match with the desired boundary condition, $\theta_m$. 


In figure \ref{fig:1} we show the results for $\theta_M$ as a function of the capillary number $Ca$. The separation of scale is kept fixed to $\lambda = \ell_s/H = 10^{-5}$ and the viscosity ratio to $\chi=1.0$. Various boundary conditions (i.e. microscopic wettabilities $\theta_m$)  are considered. In the limit of small $Ca$ the macroscopic angle is equal to the microscopic wettability $\theta_m$ but, soon after $Ca$ is increased, the interface is stretched and the macroscopic angle is decreasing. By increasing the capillary number, we clearly see that there is a  range where two solutions can exist. This is typical of fixed point structures in dynamical systems, suggesting that a bifurcation happens at a critical capillary number, $Ca_{cr}$, separating two branches, a stable branch ($\frac{d \theta_M}{d Ca}<0$)  and an unstable one ($\frac{d \theta_M}{d Ca}>0$). Above the critical capillary number no stationary solution can be found: beyond this value the interface has to evolve dynamically and liquid entrainment on the solid is expected to take place. To better stress the existence of a range of capillary numbers where two solutions are found we also show  (see right panel of figure \ref{fig:1}) the stationary interfaces corresponding to two different outer angles, $\theta_M$, with the same boundary physics, $\theta_m$. From this figure we also see that our analysis perfectly matches data presented in figure 2 of the paper by \cite{Jacqmin04} and reported in the left panel using the symbols.

\begin{figure*}
\begin{center}
\includegraphics[scale=0.26,angle=-90]{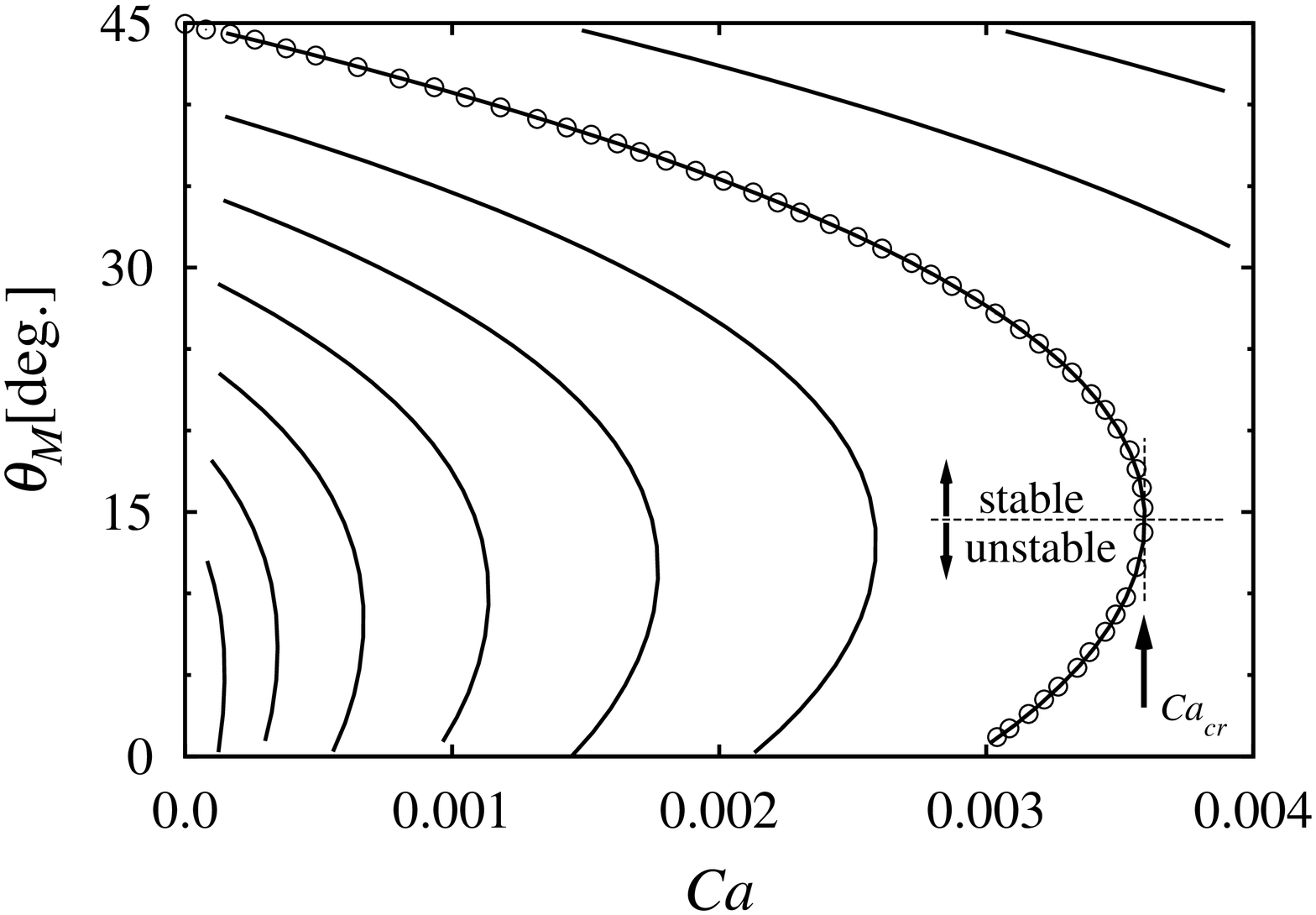}
\includegraphics[scale=0.26,angle=-90]{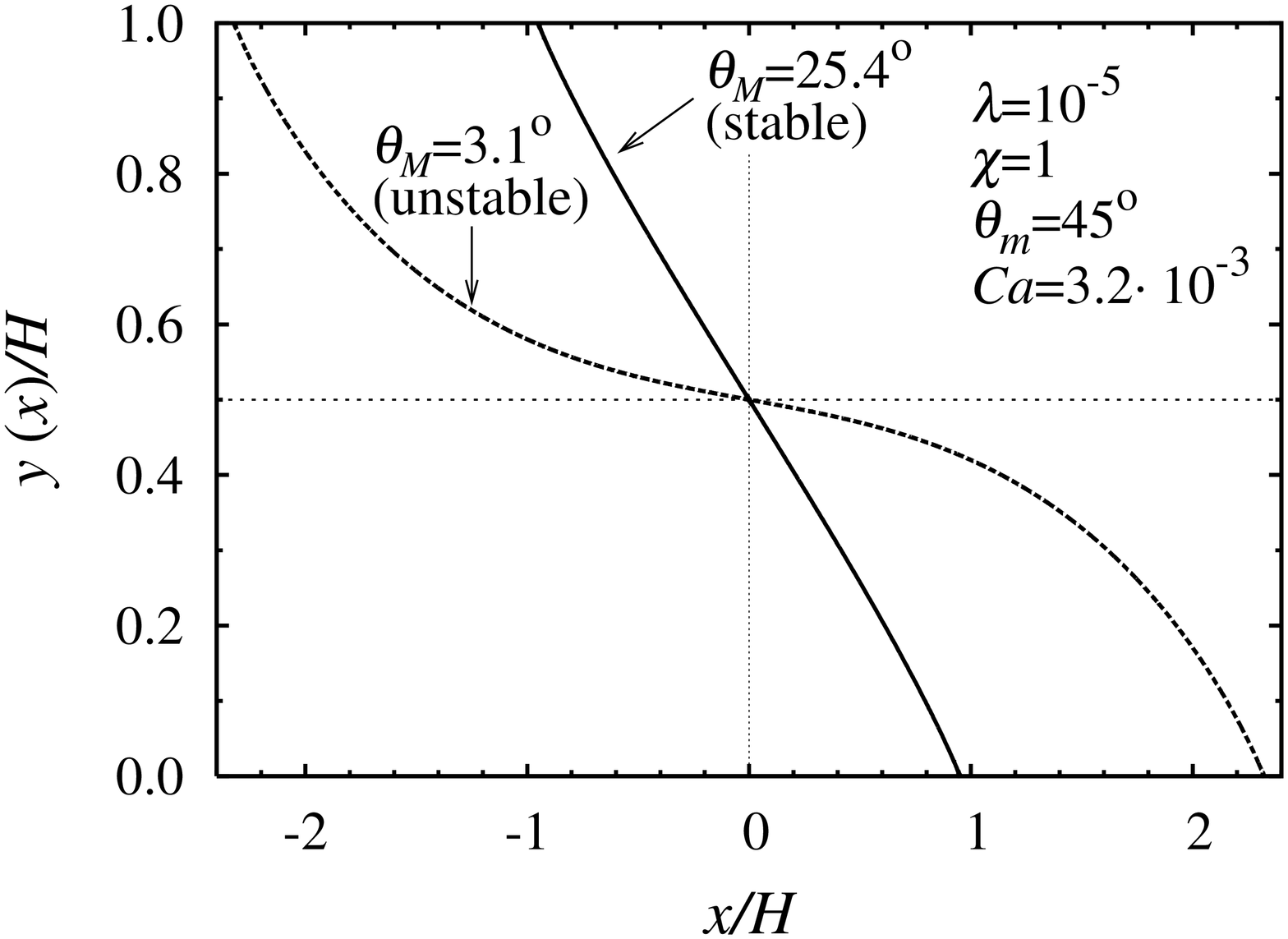}
\caption{Left: Macroscopic angle $\theta_{M}$ as a function of the
  capillary number $Ca$ for various microscopic angles  $\theta_{m}$ and for a viscous ratio $\chi=1.0$. The value of the microscopic angle corresponding to each curve  is provided by the small capillary number limit ($Ca \rightarrow 0$). The value of the separation of scale is kept constant to $\lambda=10^{-5}$. Data from Jacqmin (2004) are also reported for the case with $\theta_{m}=45^{\circ}$ (circles). On a specific curve ($\theta_{m}=45^{\circ}$), the critical capillary number, $Ca_{cr}$, above which no stationary solution exists is indicated. Right: two static profiles in the stable and unstable branch for a viscous ratio $\chi=1.0$. For the same microscopic contact angle ($\theta_m=45^{\circ}$) and separation of scale ($\lambda=10^{-5}$) we show two stationary interface profiles corresponding to the same capillary number, $Ca=0.0032$. The profile is plotted in terms of dimensionless coordinates, $y(x)/H$ and $x/H$. Results correspond to the branch shown in figure \ref{fig:1} where our results  are compared with those of Jacqmin (2004). Notice that the unstable solutions correspond to an enlarged bending of the interface close to the wall region.}
\label{fig:2}
\end{center}
\end{figure*}

\begin{figure}
\begin{center}
\includegraphics[scale=0.25,angle=-90]{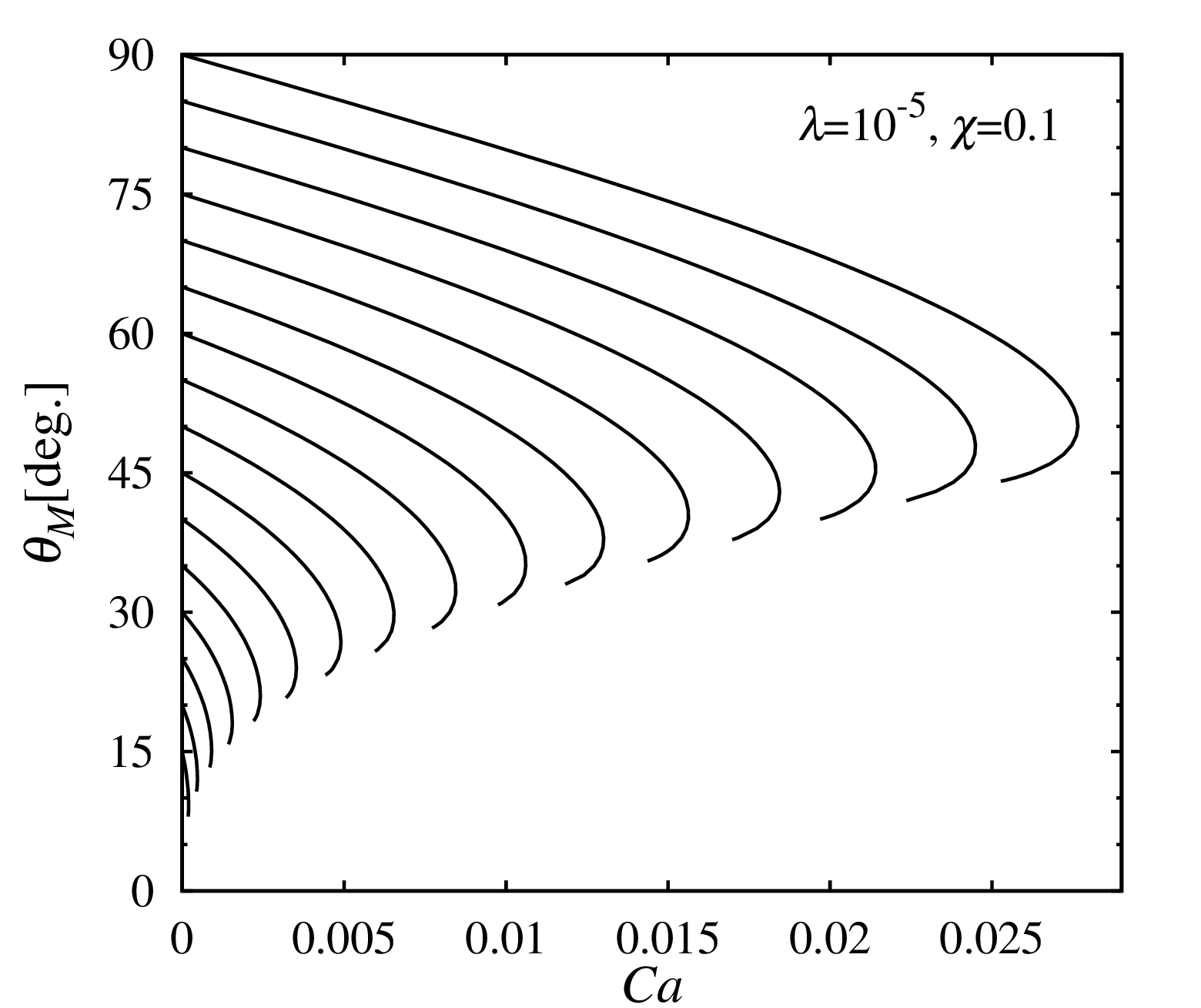}
\includegraphics[scale=0.27,angle=-90]{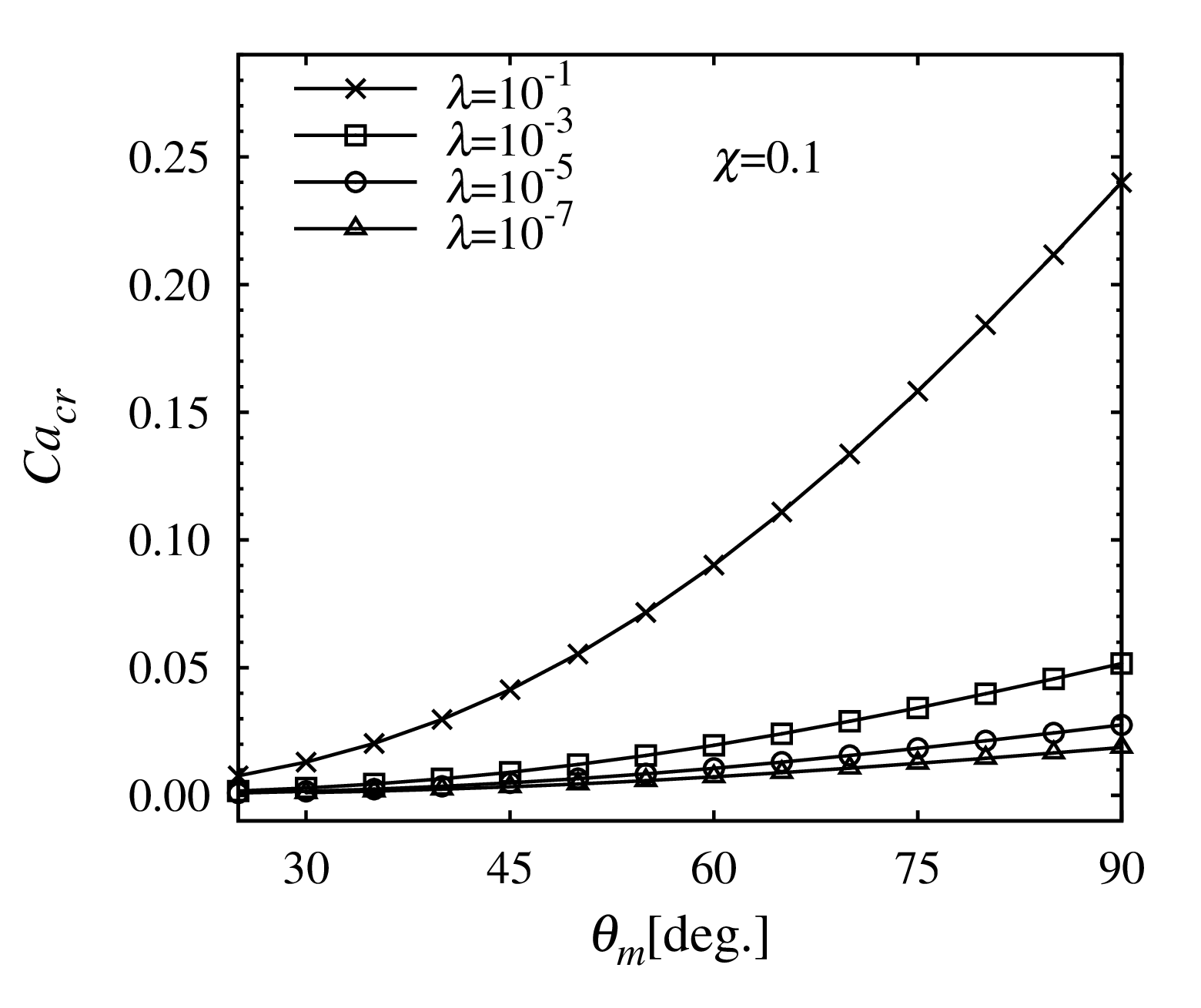}
\caption{Left: Macroscopic angle $\theta_{M}$ as a function of the
  capillary number $Ca$ for various microscopic angles
  $\theta_{m}$ and for a viscous ratio $\chi=0.1$. The value of the microscopic angle corresponding to each curve is provided by the small capillary number limit ($Ca \rightarrow 0$). The value of the separation of scale is kept constant to $\lambda=10^{-5}$ and the viscosity ratio to $\chi=1.0$. Right: the critical capillary number as a function of the microscopic angle $\theta_{m}$ for various separation of scales $\lambda$: $\lambda=10^{-1}$ (times), $\lambda=10^{-3}$ (squares), $\lambda=10^{-5}$ (circles), $\lambda=10^{-7}$ (triangles).}
\label{fig:3}
\end{center}
\end{figure}

New results are presented for the case of a different viscous ratio in figure \ref{fig:3} where we show $\theta_M$ as a function of the capillary number for a given separation of scale, $\lambda = \ell_s/H = 10^{-5}$, various microscopic wettabilities $\theta_m$, and fixed viscous ratio, $\chi=0.1$. The behavior of the unstable branch is very sensitive to the viscous ratio as well as the microscopic wettability: here we see that as soon as $\chi \neq 1$, the second (unstable) branch does not reach anymore very small values of $\theta_M$. In the right panel of the figure we show the critical capillary number as a  function of the microscopic wettability. Notice that there are remarkable variations with respect to $\lambda$ only for the larger values considered: already for $\lambda = 10^{-5}-10^{-7}$ the results are pretty stable and weakly $\lambda$ dependent. This kind of study will be important later on, when comparing with the LBE where, because of numerical limitation, one cannot ever reach separation of scales smaller then $\lambda = 10^{-2}-10^{-3}$.  Not surprisingly, figure \ref{fig:3} shows that for large scale separation the interface become more stable, i.e. $Ca_{cr}$ becomes larger.

\begin{figure}
\begin{center}
\includegraphics[scale=0.25,angle=-90]{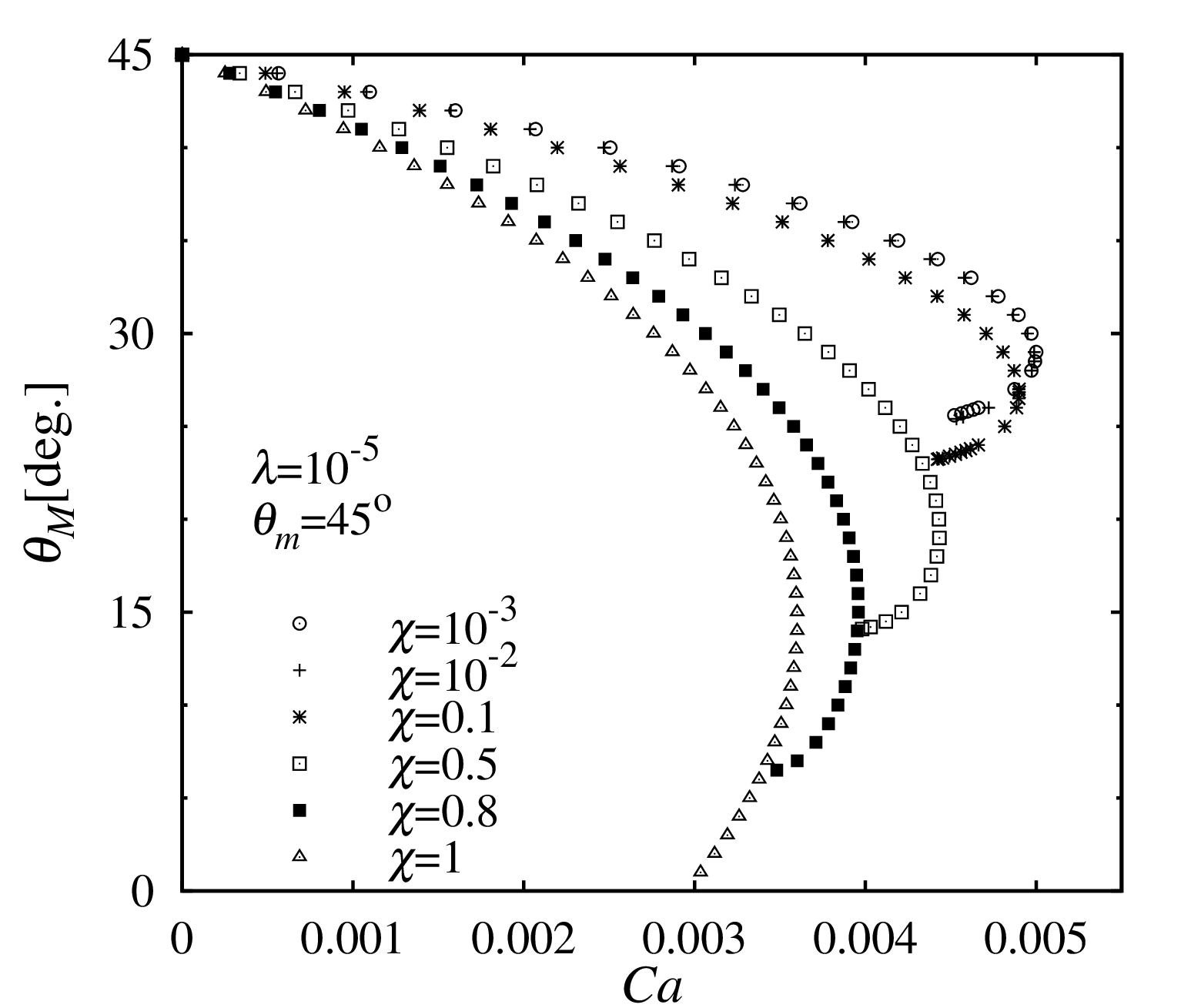}
\includegraphics[scale=0.25,angle=-90]{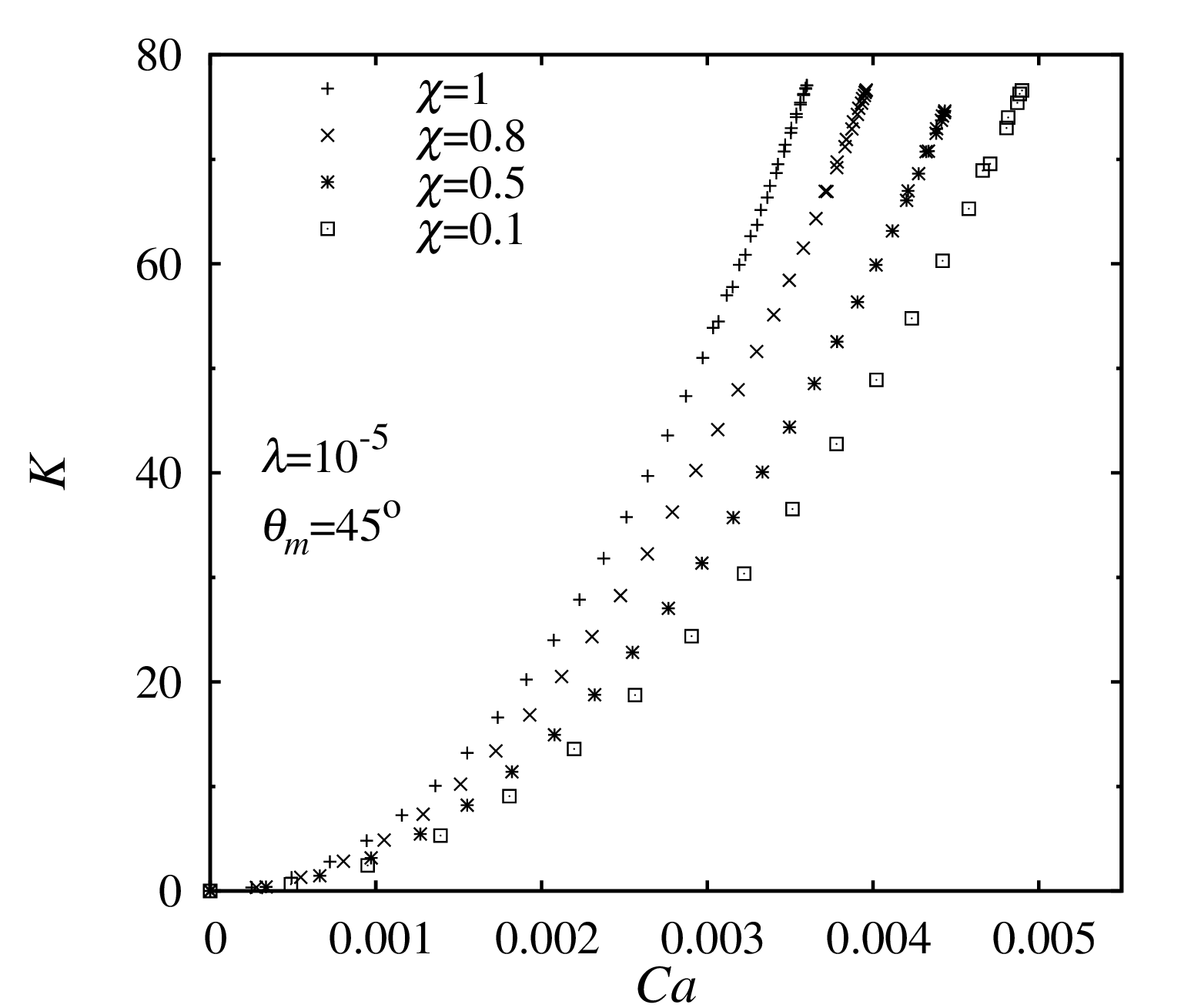}
\caption{Critical capillary number as a function of the viscous ratio. Left: for a fixed microscopic angle  ($\theta_{m}=45^{\circ}$) and separation of scale ($\lambda=10^{-5}$) we show the macroscopic angle $\theta_{M}$ as a function of the capillary number for different viscosity ratios $\chi$. Note the saturation of the curve while we are approaching values comparable with those of water and air at ordinary temperatures ($\chi \sim 0.001$). In that limit the equations are solvable exactly and we recover the usual solution for liquid-gas film. Right: We plot the integral of the squared curvature along the interface $K=\int |\partial_{xx} h(x)|^{2}dx$ as a function of the capillary number for the same microscopic wettability $\theta_m=45^{\circ}$ and $\lambda=10^{-5}$. Different viscous ratios are chosen to stress the different interface stretching.}
\label{fig:4}
\end{center}
\end{figure}

By changing the viscous ratio one can now understand the way $Ca_{cr}$ is related to $\chi$. In particular, in figure \ref{fig:4} we present results for $\theta_M$ as a function of $Ca$ for a given separation of scale $\lambda = 10^{-5}$ at changing the viscous ratio, from $\chi =1$ down to $\chi = 10^{-3}$ (a realistic value for a liquid-gas interface at ordinary temperatures).  The qualitative picture is always the same: the macroscopic angle decreases and reaches a bifurcation from which we extract the critical capillary number. We also notice that in each curve  the range of macroscopic angles associated with the two branches is decreasing with $\chi$ and also that, at decreasing $\chi$, we reach a limiting curve, i.e. results become almost independent of the viscous ratio. The fact that by decreasing the viscous ratio we increase the critical capillary number can be qualitatively understood from the equations of motion (\ref{STOKES}). When $\chi \rightarrow 0$ one of the two viscosities is getting close to zero whereas the other stays finite. By imposing  a zero viscosity in the equations of motion we tend to rule out the contribution of the pressure gradient in the less viscous fluid. Therefore, the local change in the curvature given by equation (\ref{Sder}) is triggered only by the pressure gradient of a single fluid, and we need higher capillary number to stretch it and break stationarity. All this is also complemented with the integral of the squared curvature along the profile (see right panel of figure \ref{fig:4}) for the same microscopic angle and separation of scale. The increasing stretching of the interface as a function of the viscous ratio (moving from $\chi=0$ to $\chi=1$) is reflected in this plot.

\newpage 
\subsection{Dependence on the walls wettability}

\begin{figure}
\begin{center}
\includegraphics[scale=0.5,angle=-0]{FIGURE1.eps}
\includegraphics[scale=0.5,angle=-0]{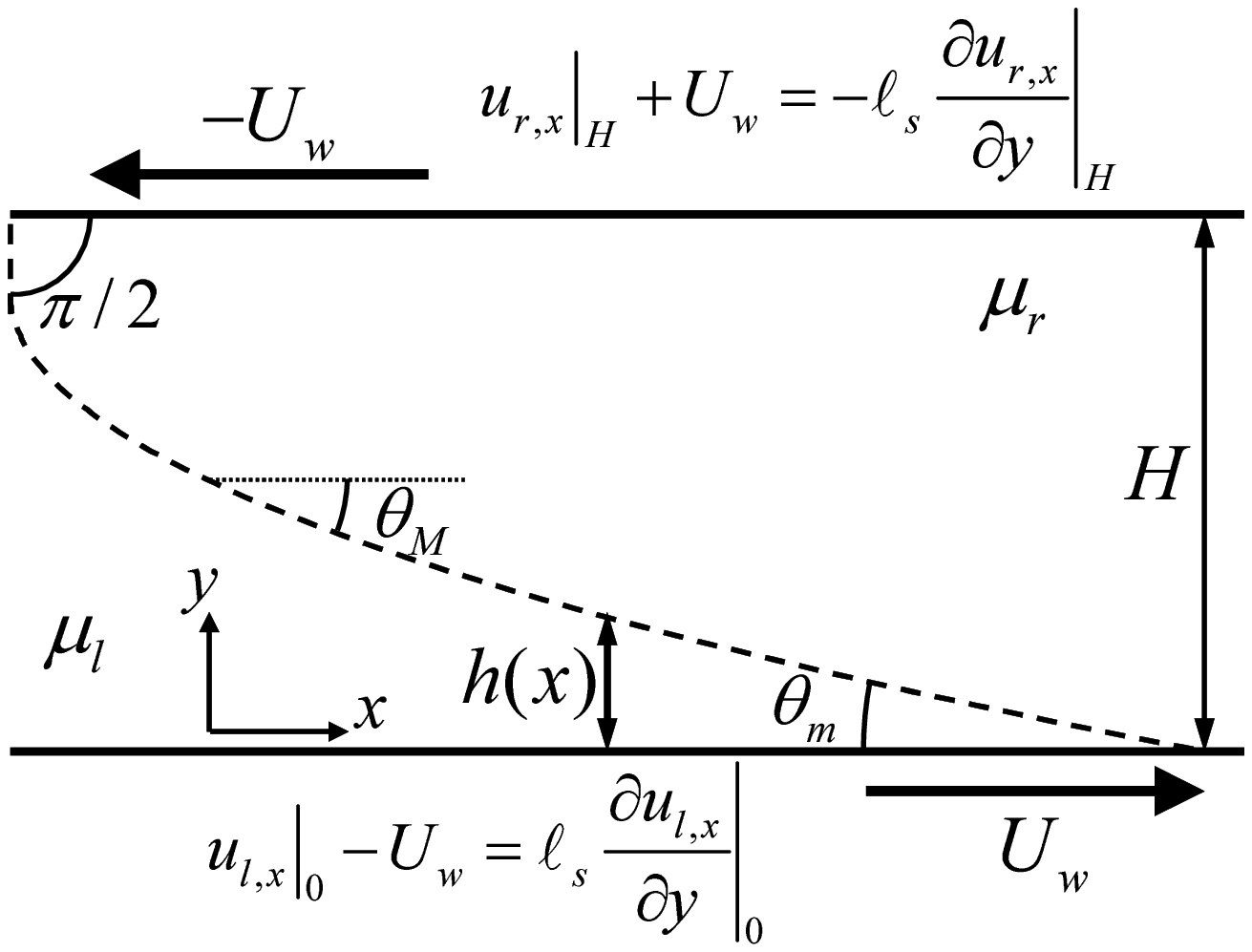}
\caption{Role of the geometry. The opposite wettability configuration (left, configuration a) is modified into another configuration (right) where the top wettability is kept fixed (configuration b). In both cases we take the macroscopic angle $\theta_M$ as the angle corresponding to the location $h(x)=H/2$.}
\label{fig:5}
\end{center}
\end{figure}

\begin{figure}
\begin{center}
\includegraphics[scale=0.4,angle=-90]{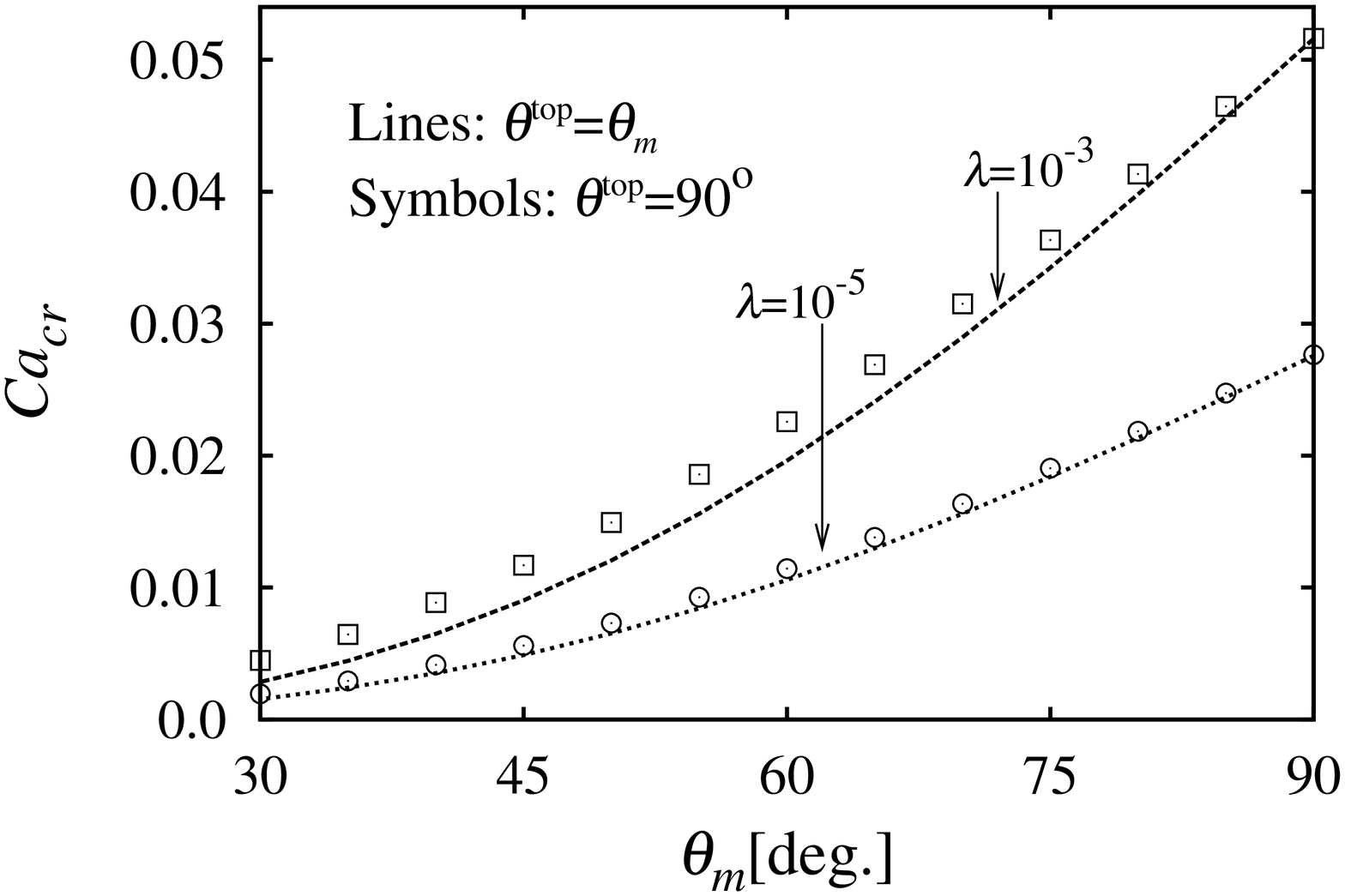}
\caption{Outer scale (non) universality. The critical capillary number is computed for the case of an opposite wettability boundary condition (see configuration a, figure \ref{fig:5}) and different separations of scales (Symbols). A similar analysis is done for configuration b (see figure \ref{fig:5}) where the top wettability has been fixed to a constant value, while changing only the lower wettability. The values of the separation of scales considered are between $\lambda=10^{-5}$ and $\lambda=10^{-3}$ and the viscous ratio is set to $\chi=0.1$.}
\label{fig:6}
\end{center}
\end{figure}

The aim of this section is to investigate the role of the wetting boundary condition at the walls. So far, we have considered the opposite wettability boundary condition as shown in figure \ref{fig:1}, where the angle the left liquid is forming with the lower boundary is the same as the angle the right fluid is forming with the upper boundary (configuration a). A possible test to understand the role of the boundary wettability is to consider another configuration (configuration b, see figure \ref{fig:5}) where the top wettability is kept fixed to a given value. The microscopic wettability is now changed only in the lower boundary. We can compute the critical capillary number in terms of $\theta_m$ and the separation of scale $\lambda$ in the same way as described in the previous sections. The separation of scales considered are between $\lambda=10^{-5}$ and $\lambda=10^{-3}$. The critical capillary number in configuration b  overestimates the case with opposite wettability (configuration a) and discrepancies are found to be enlarged when $\lambda$ is becoming larger (see figure \ref{fig:6}). The separation between the two contact line regions is indeed less pronounced when $\lambda$ is not very small, and differences can be expected to emerge in that limit. Overall, the discrepancies between the two critical capillary numbers are not that large, but this is a consequence of the small values of the separation of scale chosen.  Repeating the analysis for values of $\lambda \sim 10^{-1}$ leads to a larger discrepancy, indicating that non universalities with respect to the outer geometry become more and more pronounced by increasing $\lambda$.

\subsection{The limit of zero viscous ratio}
\label{sec:visco}

\begin{figure}
\includegraphics[scale=.32,angle=-90]{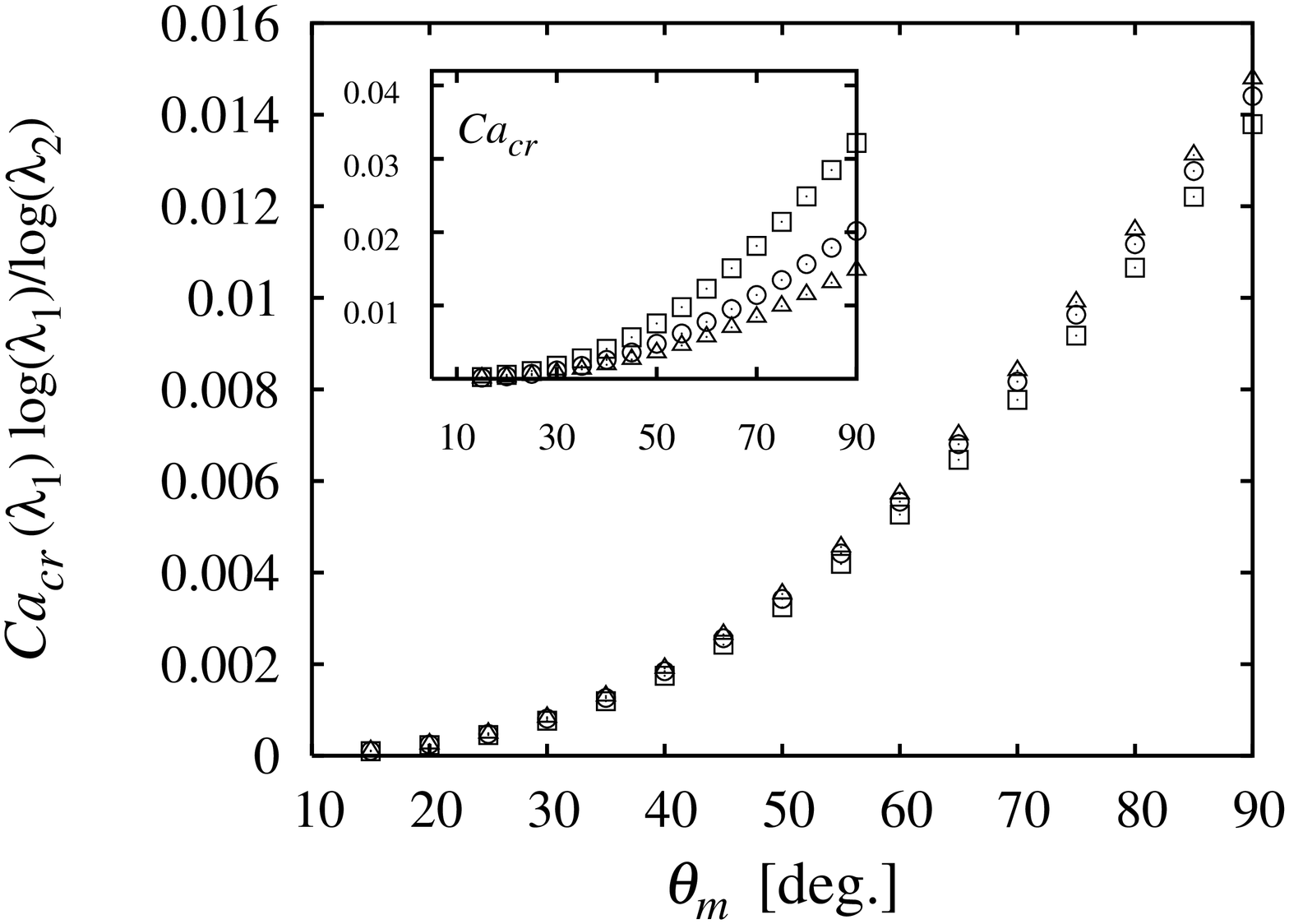}
\includegraphics[scale=.32,angle=-90]{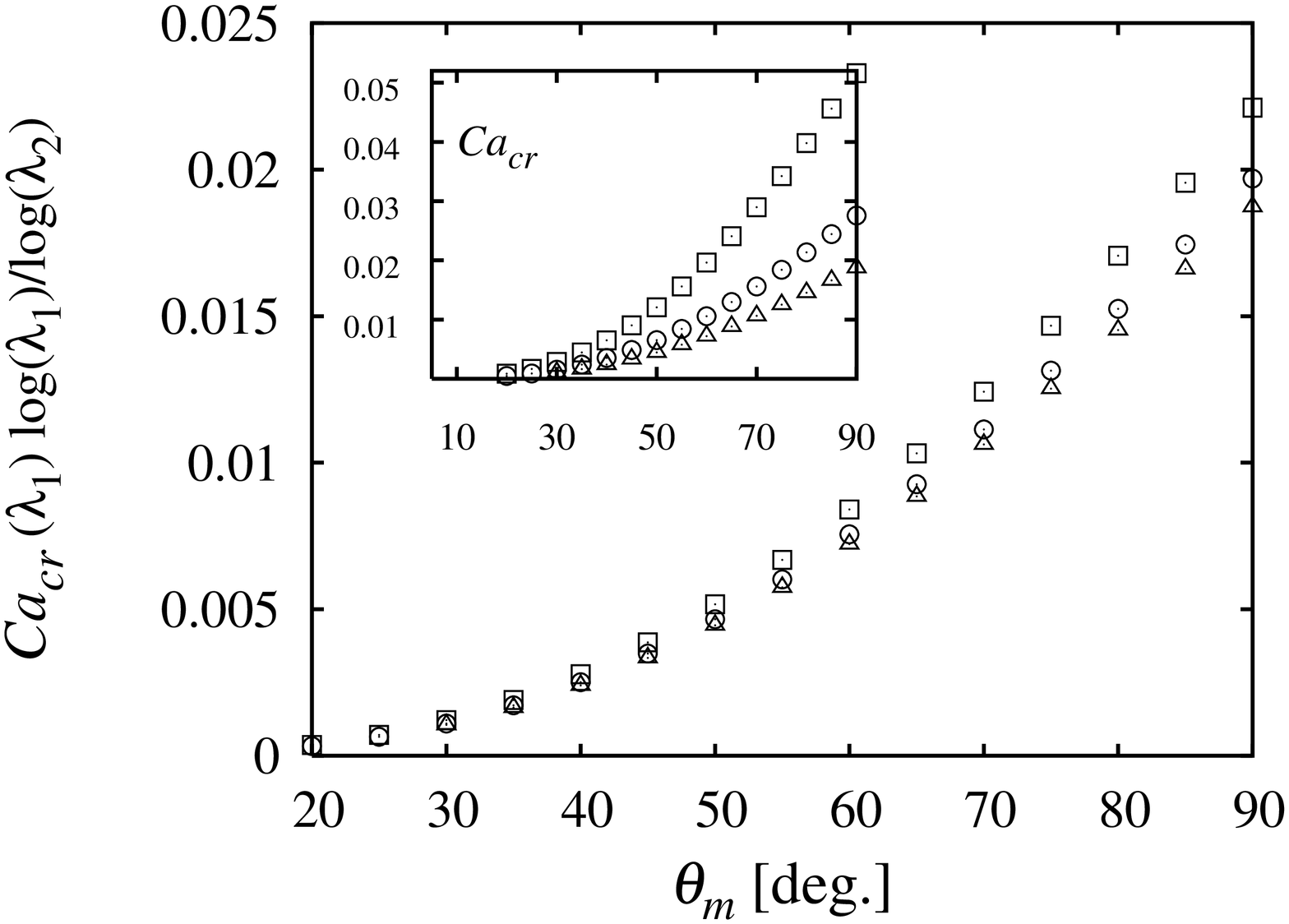}
\begin{center}
  \caption{Scaling for the critical capillary number as obtained from the full hydrodynamic calculation in the lubrication approximation. Left: the critical capillary number $Ca_{cr}$ as a function of the microscopic angle $\theta_{m}$ for $\chi=1.0$ is rescaled according to the formula predicted by theory. A separation of scale is kept fixed to $\lambda_2=10^{-7}$ (triangles). Various $\lambda_1$ are then considered: $\lambda_1=10^{-3}$ (squares), $\lambda_1=10^{-5}$ (circles). In the inset we show unscaled data (same symbols) for the critical capillary number $Ca_{cr}$ as a function of $\theta_{m}$: notice that the variability  is extensively reduced for rescaled variables (i.e. good collapse). Right: Same as left figure with a  viscosity ratio $\chi=0.1$. }
\label{fig:7}
\end{center}
\end{figure}

\begin{figure}
\begin{center}
\includegraphics[scale=.4,angle=-90]{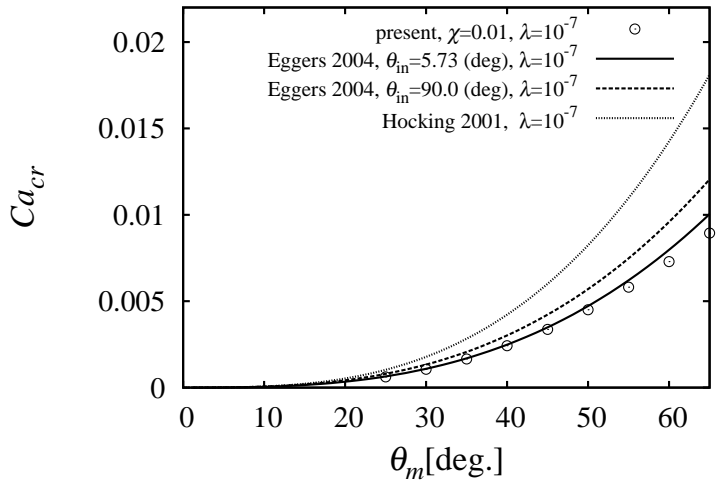}
  \caption{A comparison between our analysis and the proposed estimate for the critical capillary number given in equation (\ref{eggers}) for a plate being withdrawn from a liquid at a given speed. The angle of inclination of the plate is $\theta_{in}$ and, as it emerges from Eggers (2004), the critical capillary number is a function of both the geometry and inner physics. Typical outcomes for $\theta_{in}=5.73^{\circ}$ and $90^{\circ}$ are shown. In our case we choose the same $\lambda$ and the viscous ratio is set to a small value so as to have data already collapsed on the limiting curve for $\chi \rightarrow 0$ (figure \ref{fig:4}). Data from Hocking (2001) are also reported for the case of a  fluid in a narrow channel under the effect of gravity. }
\label{fig:8}
\end{center}
\end{figure}

We now address the properties in the limit of zero viscosity ratio, $\chi=\mu_r/\mu_l \rightarrow 0$. To do that we have to consider only the left liquid because the other component has zero viscosity and its prefluctuations ($p_{r,x}=0$ in equation (\ref{STOKES})). In this limit the lubrication equations simplify and reduce to the well known  treatment for liquid-gas interfaces (\cite{Eggers05,Eggers04,Snoeijer05,Snoeijer07}), where viscous friction is balanced with capillary effects:
\be
\frac{d^3 h}{d x^3}=-\frac{3 Ca}{h(h+\ell_s)}.
\ee
This is very similar to the inner description found in a contact line problem involving the Landau-Levich (\cite{Landau42,Derjaguin43}) geometry of a plate withdrawn at a given velocity from a liquid bath (\cite{Eggers05,Snoeijer06,Snoeijer07}). This inner solution is usually found by expanding in a power series in the capillary number  (\cite{Eggers04b,Hocking83,Voinov76}) with the final result that, away from the contact line, the cube of a macroscopic angle $\theta^3 (x) \approx \left( \frac{d h}{d x} \right)^3 $ is changing as 
\be\label{VOINOV}
\theta^3(x) \approx \theta_m^3-9 Ca \log \left( \frac{x}{\ell_s} \right).
\ee
This is basically achieved for sufficiently large arguments of $x/\ell_s$ and the related range is usually called intermediate range. When we are approaching $Ca_{cr}$, the macroscopic angle is decreasing and one can argue that the relevant scaling properties of the critical capillary number  can be captured by setting to zero the lhs of equation (\ref{VOINOV}). This implies that  $Ca_{cr}$ is rescaling like the cube of the microscopic wettability and inversely proportional to $\log \lambda^{-1}$. Therefore, for a  fixed microscopic wettability $\theta_m$, it should be expected that by changing the separation of scale there is well precise rescaling factor for the critical capillary numbers
\begin{equation}\label{SCA}
Ca_{cr}|_{\lambda_2}=Ca_{cr}|_{\lambda_1} \frac{\log \lambda_1}{\log \lambda_2}.
\end{equation}
It is interesting to check the range of applicability of the previous heuristic argument with respect to our data. In particular, in figure \ref{fig:7} we check the validity of the proposed scaling relation for two characteristic viscous ratios, $\chi=1.0$ and $\chi=0.1$. We consider the critical capillary number as a  function of the microscopic wettability $\theta_m$ and various small separation of scales $\lambda$. We compare both unscaled critical capillary numbers and scaled ones (using equation \ref{SCA}). Interestingly enough, we observe that the scaling behaviour is correct even for pretty large $\theta_m$ whereas the theory developed is based on a lubrication approximation where small titling angles are assumed (\cite{Snoeijer05}). It is interesting to observe that the scaling relation is found to hold also in the limit of equal viscosities. The term proportional to $\log \lambda$ is connected to the viscous stress divergence of the fluid at the contact line (\cite{Cox86,DeGennes86,Voinov76})  and this happens for all the values of the viscous ratio. Even if the equivalent of equation (\ref{VOINOV}) for  finite viscous ratios is different (\cite{Cox86}), the separation of scale is always appearing as $\log \lambda$ and that determines the main rescaling properties for the critical capillary number.\\
In the limit of $\chi \rightarrow 0$ it is also worthwhile to compare our estimate for the critical capillary number in the Couette cell with a similar analysis done in some recent papers by Eggers (\cite{Eggers04,Eggers05}) for the case of the Landau-Levich geometry (\cite{Landau42,Derjaguin43}) of a solid withdrawn from a liquid bath. In that case, the problem has been tackled as a multiscale problem: the inner contact line region where viscosity is balanced with capillarity must be connected with an outer region where gravity is balanced with capillary forces. An instability  occurs when the outer meniscus approaches the shape corresponding to a perfectly wetting fluid, i.e. an apparent contact angle approaching zero degrees. An investigation of the conditions under which the highly curved contact line region can be matched to the outer profile leads to the introduction of a  critical capillary number as (\cite{Eggers04,Eggers05}) 
\be\label{eggers}
Ca_{cr}=\frac{\theta^3_m}{9}\left[\log \left( \frac{Ca^{1/3}_{cr}\theta_m}{18^{1/3} \pi [\mbox{Ai}(s_{max})]^2 \lambda \theta_{in} }\right)  \right]^{-1}
\ee
where $\theta_{in}$ is the angle of inclination of the plate with respect to the liquid bath, Ai is the Airi function and $s_{max}=-1.0188...$ is the point where the Airi function assumes its maximum. The setup is clearly different from our Couette cell but neverthless is useful to compare both results to understand the role of the geometry in determining the  critical capillary number. We choose the same separation of scale, $\lambda=10^{-7}$, and a small viscous ratio, $\chi=0.01$, in the Couette cell. This is done to be in the region where we observe the collapse of data, as discussed in figure \ref{fig:4}. We use two characteristic angles of inclination $\theta_{in}$ in (\ref{eggers}) ($\theta_{in}=5.73^{\circ}$ and $\theta_{in}=90^{\circ}$) and the results are plotted in figure \ref{fig:8}. The order of magnitude of the different critical capillary numbers is comparable and data share the same scaling properties with respect to $\theta_m$ ($Ca_{cr} \sim \theta^3_{m}$). Anyhow, the presence of an overall prefactor dependent on the geometry is clearly visible from the figure. For the sake of completeness, we have also considered a recent  prediction of \cite{Hocking01} for fluids in narrow channels. The geometry is that of a  fluid confined between two parallel plates at distance $2d$ under the effect of gravity. The main difference with the previous cases is that the speed of withdrawal is not an additional parameter but is fixed in terms of the bulk forcing, i.e. gravity. After introducing a  Bond number, $Bo=\frac{\rho_l g d^2}{\sigma}$, with $\rho_l$ the liquid density and $g$ the gravity, one can find the connection between $Bo$ and the capillary number to be $Ca=Bo/3$. In this notation $\mu_l$ is the liquid viscosity and $U$ is the average speed of motion due to gravity. A critical value in the Bond number (equation (17) in Hocking(2001)) is then translated into a critical capillary number as
\be\label{Hocking}
Ca_{cr} \approx \frac{0.6}{3 \log(\lambda^{-1})} \theta_m^3.
\ee 
The plot of (\ref{Hocking}) is also considered in figure \ref{fig:8} with the same $\lambda$ of the previous estimates. This prediction turns out to overestimate all previous curves.

\newpage

\section{Numerical Approach: diffuse interface Methods and Lattice Boltzmann equation (LBE)}
\label{sec:lbe}

In this section we will use  numerical simulations to further investigate the problem previously discussed. An approach based on first principles to address the  inner physics of contact line problems is molecular dynamics (MD). These simulations, which usually involve Lennard-Jones liquids and several thousand molecules, appeared to exhibit continuum behavior at the macroscopic level (\cite{Thompson89,Koplik89,Denniston01,Barrat99}). All the results of these molecular dynamics simulations offer insight especially in the region close to the contact line, pointing to the breakdown of the no-slip boundary condition at very small distances as a possible explanation of  contact line motion (\cite{Thompson89,Barrat99}). Alternative explanations of contact line motion exist and they do not rely on the breakdown of  the no slip boundary condition. As was noticed by \cite{Seppecher96}, a sharp interface model may be questioned. It is itself an approximation which may not be valid in the vicinity of the contact line. Then one may use molecular simulations or a continuum model able to describe the interface as a diffuse layer, i.e. a layer of finite thickness. The curvature of the interface near the contact line leads to mass transport across the interface, thus releasing the viscous singularity. Similar analysis in binary fluid systems (\cite{Jacqmin00,Chen00}) showed that diffusive transport of the fluid leads to effective slip at the contact line.  Also mesoscopic diffuse interface models  for two-phase flows based on the lattice Boltzmann equation (\cite{Gladrow,Saurobook})  were applied to the simulation of contact line motion and related problems (\cite{Yeomans04a,Yeomans04b,Latva07,Jia06,Kwok06}). In particular we further elaborate along these lines using mesoscopic diffuse interface models where multi-phase physics is induced by using a pseudo-potential approach originally developed by \cite{SC93,SC94}, hereafter SC. 


\subsection{The Lattice Boltzmann Equation for non ideal fluids}

We start from the usual lattice Boltzmann equation with a single-time
relaxation (\cite{LBGK,Gladrow,Saurobook}): 
\be\label{eq:LB}
f_{l}(\bm{x}+\bm{c}_{l}\Delta t,t+\Delta
t)-f_l(\bm{x},t)=-\frac{\Delta t}{\tau}\left(
  f_{l}(\bm{x},t)-f_{l}^{(eq)}(\rho,\rho {\bm u}) \right) +F_l 
\ee
where $f_l(\bm{x},t)$ is the kinetic probability density function associated with a mesoscopic velocity $\bm{c}_{l}$ (there is a  discrete set of velocities), $\tau$ is a mean collision time (with $\Delta t$ a time lapse), $f^{(eq)}_{l}(\rho,\rho {\bm u})$ the equilibrium distribution, corresponding to the Maxwellian distribution in the continuum limit and $F_l$ represents a general forcing term whose role will be discussed later in the framework of inter-molecular interactions.  From the kinetic distributions we can define macroscopic density and momentum fields as (\cite{Gladrow,Saurobook}): 
\be\label{Momentum}
\rho(\x)=\sum_{l} f_{l}(\x); \qquad  \rho
{\bm u}(\x)=\sum_{l}{\bm c}_{l}f_{l}(\x).  
\ee 
For technical details and numerical simulations we shall refer to the nine-speed, two-dimensional $2DQ9$ model (\cite{Gladrow}), often used due to its numerical robustness. The equilibrium distribution in the lattice Boltzmann equations is obtained via a low Mach number expansion of the continuum Maxwellian (\cite{Gladrow,Saurobook})
\be\label{EQUILIBRIUM}
f^{(eq)}_{l}={w}^{(eq)}_{l}\left[ \rho+\frac{c^{i}_{l} \rho
    u_{i}}{{c}^{2}_{s}}+
  \frac{(c^{i}_{l}c^{j}_{l}-{c}^{2}_{s}\delta_{ij})}{2 {c}^{4}_{s}}
  \rho u_{i}u_{j} \right] 
\ee 
where ${{c}}^{2}_{s}=1/3$ and $i=1,2=x,y$ runs over spatial dimensions.  The weights ${w}^{(eq)}_{l}$ are chosen such as to enforce isotropy up to fourth order tensor in the lattice (\cite{Gladrow,Saurobook}). From the equilibrium distribution and the symmetry properties of ${\bm c}_l$, it immediately follows (\cite{Gladrow}) the kinetic second order tensor of the equilibrium distribution:
$$
\sum_{l}f^{(eq)}_{l}{c}^{i}_{l}{c}^{j}_{l}=\delta_{ij}({{c}}^{2}_{s}\rho)+ \rho u_{i}u_{j},
$$
where, in the first term of the rhs, we recognize the  ideal-gas pressure tensor: $ P_{ij} =\delta_{ij}(c_s^2 \rho).$  In order to study non-ideal effects we need to supplement the previous description with an inter-particle forcing. This is done by choosing a suitable $F_l$ in (\ref{eq:LB}). In the original SC model, the bulk inter-particle interaction is proportional to a free parameter (the ratio of potential to thermal energy), ${\cal G}_{b}$, entering the equation for the momentum balance: 
\be\label{forcing} 
F_{i}=\sum_l F_l {c}^{i}_l=-{\cal G}_{b}
c^{2}_{s}\sum_l w(|{\bm c}_{l}|^2) \psi(\x,t) \psi (\x+{\bm c}_l\Delta
t,t) {c}^{i}_l 
\ee 
being $w(|{\bm c}_{l}|^2)$ the static weights and $\psi(\x,t)=\psi(\rho(\x,t)$ the (pseudo) potential function which describes the fluid-fluid interactions triggered by inhomogeneities of the density profile (\cite{Gladrow,Sbragaglia07}). We shall refer to the pseudo-potential used in the original SC work, namely 
\be\label{PSI} 
\psi(\x,t)=(1-exp(-\rho({\bm x},t))).  
\ee 
Note that this reduces to the correct form $\psi \rightarrow \rho$ in the limit $\rho
\ll 1$, whereas at high density ($\rho \gg 1$), it shows a saturation. This latter is crucial to prevent density collapse of the high-density phases (note that the SC potential is purely attractive, so that a mechanism stabilizing the high-density phase is mandatory to prevent density collapse).  In principle, other functional forms may be investigated, sometimes with impressive enhancement of the density ratios supported by the model (\cite{Yuan06}).\\
In order to understand the corrections to the ideal-state equation  induced by the pseudo-potential, we need to define a consistent pressure tensor, $P_{ij}$, for the macroscopic variables:
\be
\label{eq:cons}
\partial_j P_{ij} = -F_i +\partial_i (c_s^2 \rho).  
\ee 
Upon Taylor expanding the forcing term and assuming hereafter $\Delta t=1$, we obtain 
\be
\label{taylor}
F_{i}=-{\cal G}_{b}\psi \partial_{i}\psi - \frac{{\cal G}_{b}}{2}\psi \partial_{i} \Delta \psi 
\ee
which is correctly translated (\cite{Benzi06,He02}) into 
\be\label{TENSOREold} 
P_{ij}=\left( c^{2}_{s}\rho+{\G}_{b}\frac{c_{s}^{2}}{2}\psi^{2}+{\G}_{b}\frac{c^{4}_{s}}{4}|{\bf \nabla}\psi|^{2} +{\G}_{b} \frac{c^{4}_{s}}{2}\psi \Delta \psi \right) \delta_{ij}- \frac{1}{2}{{\G}_{b} {{c}}^{4}_{s}}\partial_{i} \psi \partial_{j}\psi.
\ee
The evolution scheme (\ref{eq:LB}) together with the inter-particle force (\ref{forcing}) approximates the  following diffuse interface equations for the density and momentum fields (\ref{Momentum}):
\be
\partial_{t} \rho+\partial_{j} (\rho u_{j})=0
\ee
\be
\partial_{t}(\rho u_{i})+ u_{j} \partial_{j} (\rho u_{i})=-\partial_{j}P_{ij}+\partial_{j} \Pi_{ij} 
\ee 
with $\Pi_{ij}$ the usual viscous stress tensor (\cite{Gladrow,Saurobook,Yeomans04a,Yeomans04b}).  Capillary effects, mimicking non-trivial interactions with the wall, can be incorporated in this approach by using a suitable wall function (\cite{Sbragaglia06,Sbragaglia07}).\\ 
A disadvantage of the SC formulation is that, existing only one free parameter to tune ($\G_{b}$), one cannot change independently the density ratio and the surface tension. Here we use a recent extension of the SC model that overcomes this limitation, by introducing first and
second neighbors coupling in the pseudo-potential (\cite{Sbragaglia07}):
\be\label{F2}
F_{i}=-c_s^2 \sum_l w(|{\bm c}_{l}|^2) \psi(\x,t) [{\G}_{1} \psi
(\x+{\bm c}_l,t)+{\G}_{2} \psi (\x+2{\bm c}_l,t)] {c}^{i}_l.
\ee
This coupling leads to the following Pressure tensor in the continuum limit: 
\be\label{TENSORE}
P_{ij}=\left(
  c^{2}_{s}\rho+{\G}_{1}\frac{c_{s}^{2}}{2}\psi^{2}+{\G}_{2}\frac{c^{4}_{s}}{4}|{\bf
    \nabla}\psi|^{2} +{\G}_{2} \frac{c^{4}_{s}}{2}\psi \Delta \psi
\right) \delta_{ij}- \frac{1}{2}{{\G}_{2} {{c}}^{4}_{s}}\partial_{i}
\psi \partial_{j}\psi. 
\ee 
Furthermore, in order to reduce the importance of spurious currents at interfaces (\cite{Sbragaglia07,Shan06,Lee06,Wagner03,Cristea03,Yuan06}) one can also introduce the non-ideal terms directly into the equilibrium distribution as explained in some recent papers (\cite{Yeomans04a,Yeomans04b}). With this strategy we impose the desired pressure tensor instead of writing explicitly the forcing term (\ref{F2}). We can thus use ${\G}_{1}$ and ${\G}_{2}$ to vary density ratio and width of the interface (i.e. surface tension) independently. As a matter of fact, the use of Van der Waals pressure tensors (\cite{Yeomans04a,Yeomans04b}), may limit seriously the density ratios achieved in the simulations.

\subsection{Stationary Contact Line description: Cox (1986) {\it vs} LBE}

\begin{figure}
\includegraphics[scale=.32,angle=-90]{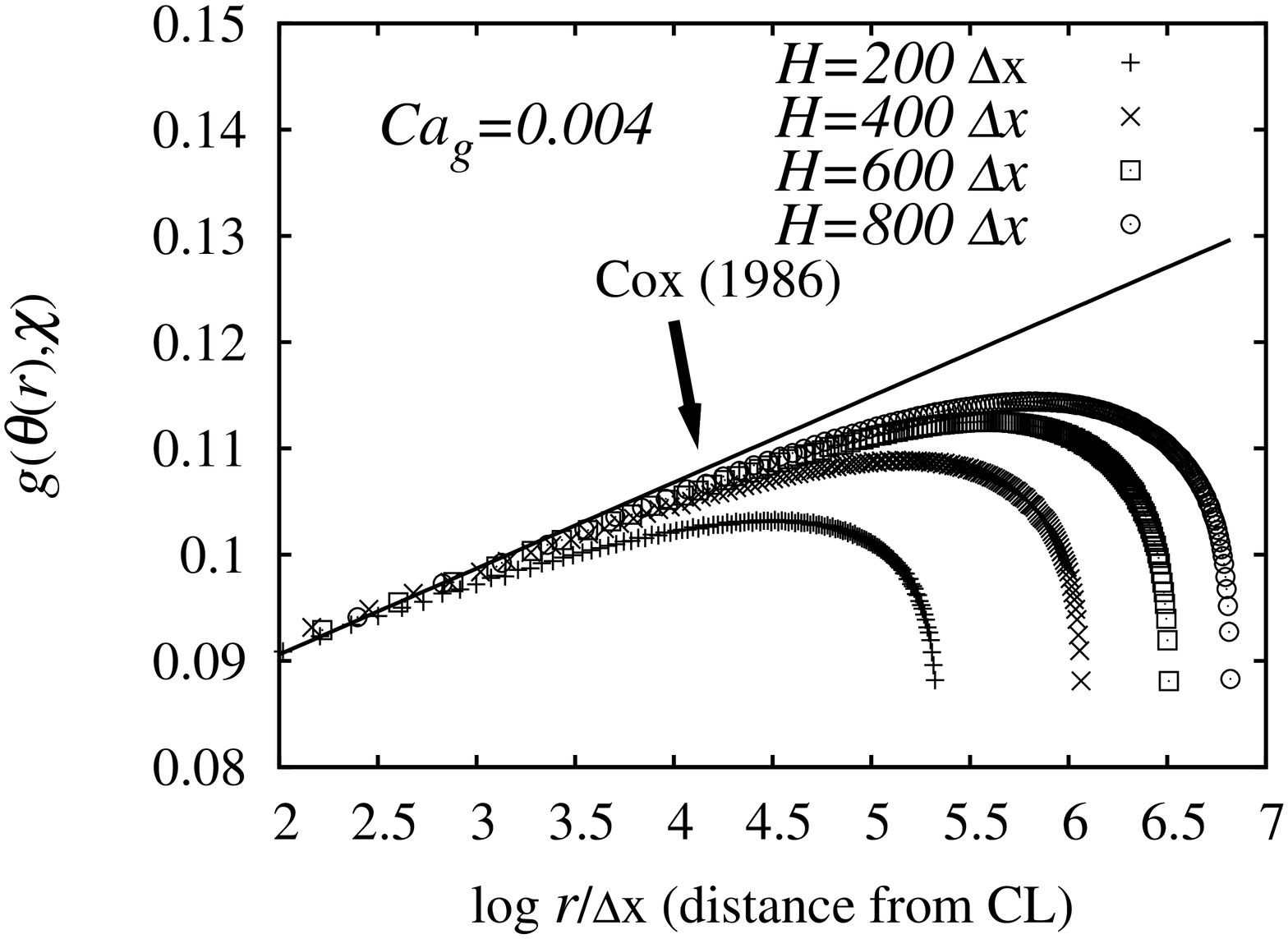}
\includegraphics[scale=.32,angle=-90]{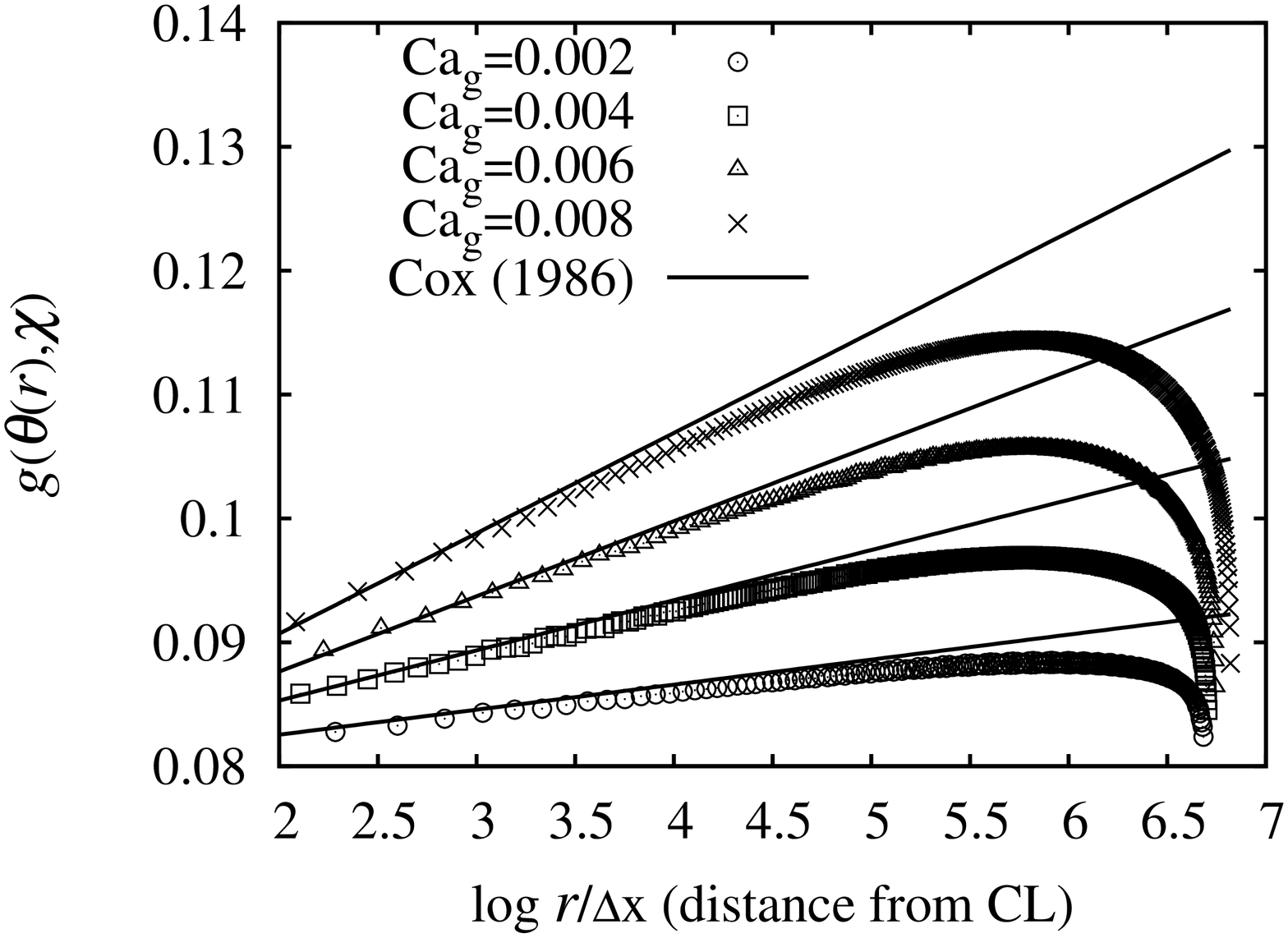}
\begin{center}
  \caption{The comparison between numerical simulations and the scaling relations for the intermediate region as predicted by Cox (1986). Left: The scaling for $g(\theta,\chi)$ as a  function of the distance from the contact line $r$. For a fixed capillary number in the gas phase, $Ca_g$, fixed viscosity ratio, $\chi$, and fixed slip length, the sharp interface  theory predicts a linear scaling of the function $g(\theta,\chi)$ (see equation (\ref{eq:COX})) with respect to $\log r$. The development of the intermediate region is shown for various numerical resolutions approaching the sharp interface limit, where theory and numerics become asymptotically comparable. In the intermediate region the slope is the one predicted by theory, i.e. equal to $Ca_{g}=\frac{\mu_r U_w}{\sigma}$. Right: the same as left figure with different capillary numbers and a given resolution in the vertical direction.}
\label{fig:9}
\end{center}
\end{figure}

In this subsection we explore and investigate the conditions under which our diffuse interface models converge to the predictions of sharp interface hydrodynamics. As we have already seen and discussed in equation (\ref{VOINOV}), we should expect an intermediate range with a prediction of scaling for the slope of the interface in terms of the capillary number and the microscopic slip length. A  further extension for that prediction of scaling for two fluids with finite viscous ratio has been provided by \cite{Cox86}. This scaling relation translates (see figure 1 in Cox (1986)) in our coordinates (see figure \ref{fig:1}) as
\be\label{eq:COX}
g(\theta(r),\chi)=g(\theta(0),\chi)+Ca_{g} \log (r/\ell_s)
\ee
\be
g(\theta,\chi)=\int_{0}^{\theta} \frac{1}{f(x,\chi)} dx
\ee
and
\be
f(x,\chi)=\frac{2 \sin x \left[\chi^{-2}(x^2-\sin^2 x)+2 \chi^{-1}(x(\pi-x)+\sin ^2 x)+((\pi-x)^2-\sin^2 x) \right]}{\chi^{-1}(x^2-\sin^2 x)((\pi-x)+\sin x \cos x)+((\pi-x)^2-\sin^2 x)(x-\sin x \cos x)}
\ee
where we have defined $Ca_g=\frac{\mu_r U_w}{\sigma}$ as the capillary number estimated in the less viscous fluid, $r$ as the distance from the stationary contact line and the angle $\theta(r)$ consistently with the notation of figure 1 in the paper of \cite{Cox86}. This is a pure sharp interface treatment of the moving contact line involving a  finite viscous ratio and the length scale $\ell_s$ is associated with microscopic slip motion.  This turns out to be extremely useful to compare with our simulations where it is difficult to reach very small viscous ratios. We would expect that in our diffuse interface methods, when we look at the contact line on scales much larger than the interface width, $\xi$, we should recover a quantitative matching with  the sharp interface prediction (\ref{eq:COX}). The length scale $\ell_s$ is now associated with the effective slip generated by the diffuse interface mechanisms (\cite{Seppecher96,Jacqmin00,Pismen00}). This study opens the possibility to test the universality in the intermediate region independently of the inner physics mechanisms. In our numerical simulations we fix the viscous ratio to be equal to $\chi=0.1$ and the associated density ratio is chosen so that the interface width $\xi$ is of the order of some grid points \footnote{One would need a systematic procedure to give $\xi$. A possible way is to define $\xi$ in terms of those points where the local gradient exceeds a given threshold. In any case the width $\xi$ does not exceeds some units and it is reassuring to assume $\xi/\Delta x=4 \pm 1$ for the case of $\chi=0.1$.}. The wetting properties are chosen as explained in some recent studies (\cite{Benzi06,Sbragaglia07}). In particular, for the present set of simulations, we use a neutral wetting boundary condition ($\theta(0)=\pi/2$ in (\ref{eq:COX})). To do that we assume that the pseudopotential of the lattice Boltzmann description (\ref{PSI}) achieves a neutral stress at the boundaries ($\partial_n \psi=0$, with $n$ the normal with respect to the wall). With a fixed $Ca_g$, interface width $\xi$ and viscous ratio $\chi=0.1$ we then carry out numerical simulations in our Couette cell by enlarging the distance between the walls: $H=200 \Delta x, 400 \Delta x, 600 \Delta x, 800 \Delta x $ (see figure {\ref{fig:9}} left panel), with $\Delta x$ the lattice spacing. Half of the computational domain has then been initialized with the left, more viscous fluid and rotational boundary conditions are implemented at the inlet and outlet as discussed by \cite{Yeomans04b}. No slip boundary conditions are implemented exactly, i.e. the mechanism of releasing the contact line singularity is found in compressibility effects due to the diffuse nature of the model. The results of the numerical simulations (see figure \ref{fig:9}) clearly show that by enlarging the resolution, i.e. increasing the ratio of the outer scale $H$ with respect to the interface width, we correctly approach the linear scaling behaviour of $g(\theta(r),\chi)$ with respect to $\log r$. The slope is correctly given by $Ca_{g}$ in (\ref{eq:COX}). The linearity with respect to $Ca_{g}$ is further checked in the right panel of figure \ref{fig:9} where we keep fixed $H$ and simply change the capillary number by changing the velocity at the wall $U_w$.

\subsection{Full Hydrodynamic Calculations: Lubrication Theory {\it vs} LBE}

\begin{figure}
\includegraphics[scale=.32,angle=-90]{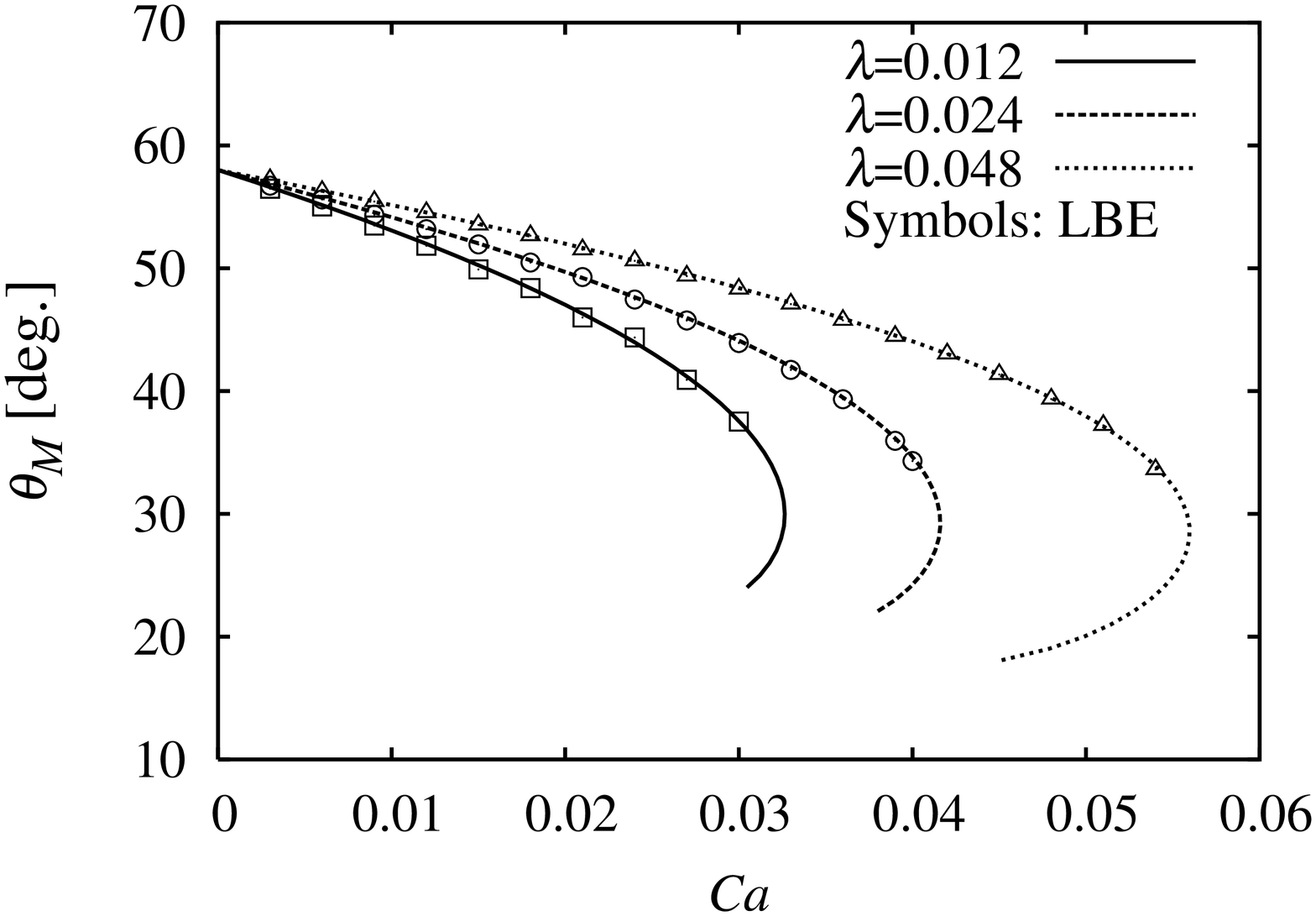}
\includegraphics[scale=.32,angle=-90]{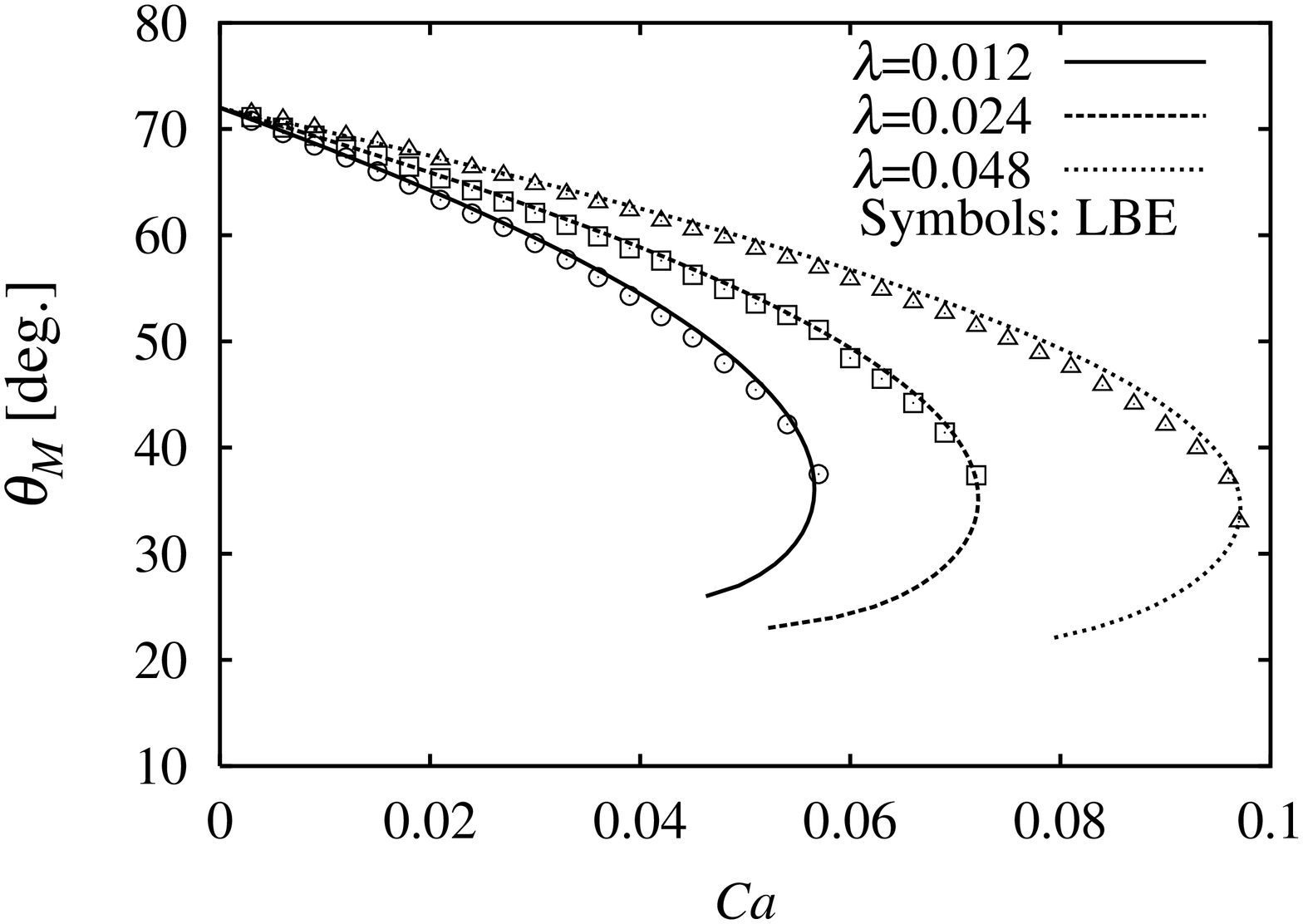}
\begin{center}
  \caption{The comparison between theory and numerical simulations for the macroscopic angle $\theta_M$ as a function of the capillary number $Ca$ on the stable branch. Left: for  a fixed microscopic angle $\theta=58^{\circ}$,  three different scale separations $\lambda$  are used: $\lambda=0.048$ (triangles), $\lambda=0.024$ (circles), $\lambda=0.012$ (squares). In the numerics the different scale separations are obtained by fixing the interface width in the diffuse interface approach and using different heights  $H$. Left: same as right figure for a microscopic contact angle $\theta=72^{\circ}$.}
\label{fig:10}
\end{center}
\end{figure}

In the previous subsection we have quantitatively matched the diffuse interface methods with the universal scaling properties in the intermediate range, as predicted by \cite{Cox86}. To achieve a good comparison, a lattice Boltzmann separation of scale, $\lambda_{LBE}=\xi/H$, must be chosen as a small scale parameter, i.e. the scales of observation must be larger with respect to the inner length scale specified by the interface width $\xi$. Clearly, $\lambda_{LBE}$ plays the role of the separation of scale in the diffuse interface description. In particular, we can investigate the possibility of a quantitative matching with the prediction coming from the full hydrodynamical calculation developed in the first part of the paper. This would better explore the comparison between the two descriptions (sharp {\it vs} diffuse) on all length scales, and not simply in the intermediate range as done in the previous subsection. To do that we choose the same viscous ratio as before ($\chi=0.1$) and we fix the distance between the walls to be $H=100 \Delta x$. The microscopic wettability is chosen to be equal to $\theta_m=58^{\circ}$ and then we vary the capillary number $Ca$. For each value of $Ca$ we reach the stationary state and collect the corresponding value of the angle in the center of the cell ($\theta_M$). To extract the angle we do contour plots of the density field on the interface at the level $\rho_{av}=\frac{\rho_g+\rho_l}{2}$, with $\rho_{g,l}$ the gas/liquid density respectively. The corresponding plot of $\theta_{M}$ {\it vs} $Ca$ is shown in the left panel of figure \ref{fig:10}. Notice that when we reach a critical value of the capillary number, no stationary interface is observed above that value. This is precisely the same behaviour as predicted by theory. In particular, we can find the correct value of $\lambda$ in the sharp interface treatment of the Couette cell able to fit correctly the numerical prediction. It is found that the correct value of $\lambda$ is approximately $\lambda=0.024$, which is of the order of $\lambda_{LBE}$. We can also repeat the whole experiment with the viscous ratio and interface width fixed to the same values, but doubling or reducing by a factor $2$ the resolution in between the walls. We would expect and we observe (see figure \ref{fig:10}) that the corresponding numerical data match with the theoretical prediction obtained  by doubling or reducing by a factor $2$ the previously used $\lambda$. The same study is also done with a different microscopic wettability ($\theta_m=72^{\circ}$, see right panel of figure \ref{fig:10}). Overall, we observe a  good agreement between the two approaches.

\begin{figure}
\includegraphics[scale=.32,angle=-90]{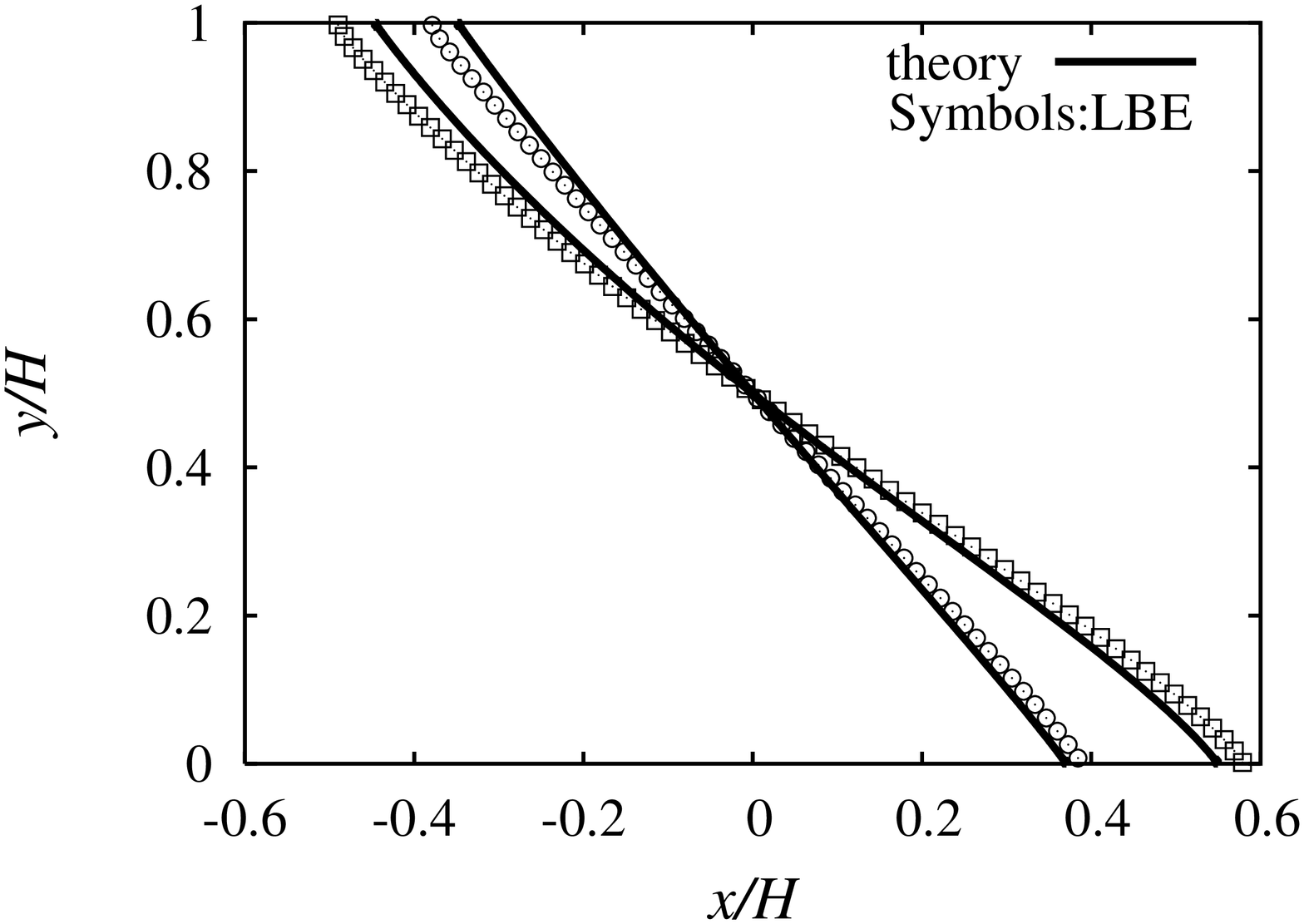}
\includegraphics[scale=.32,angle=-90]{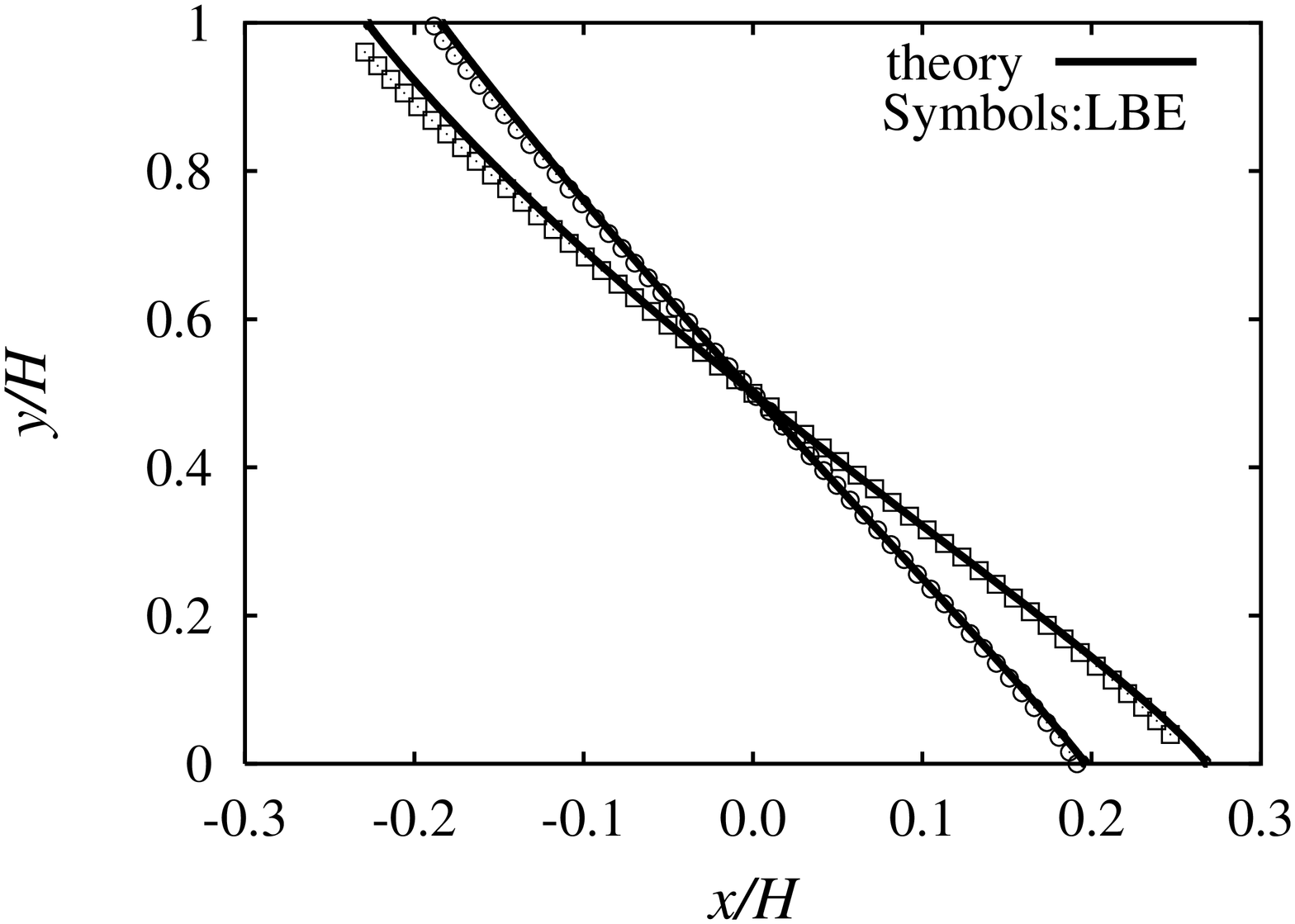}
\begin{center}
  \caption{The static interface shape for a given capillary number and microscopic angle $\theta_{m}$. Left: for $\theta_{m}=58^{\circ}$ we show the results of the numerical simulations for two different capillary numbers: $Ca=0.033$ (squares), $Ca=0.012$ (circles). The corresponding theoretical prediction is shown (lines). The horizontal and vertical coordinate have been made dimensionless with respect to the system's height $H$. The separation of scale $\lambda$  in the theory is $\lambda=0.024$ that is the correct one to reproduce the diffuse interface results for the macroscopic angle $\theta_M$ as a  function of the capillary number $Ca$ (see figure \ref{fig:10}).  Right: same as left for $\theta_{m}=72^{\circ}$  }
\label{fig:11}
\end{center}
\end{figure}

In figure \ref{fig:11} we compare stationary interface profiles for a  fixed viscous ratio, $\chi=0.1$, and two characteristic capillary numbers, $Ca=0.022$ and $Ca=0.033$. We choose $H=100 \Delta x$ and the interface width $\xi$ is kept fixed. The corresponding prediction from the theory has been produced with the value of $\lambda$ able to match the corresponding numerics in the plot of $\theta_M$ as a  function of $Ca$ (see middle plots of figure \ref{fig:10}). It should be noticed that we match correctly the bulk regions, whereas small discrepancies emerge close to the boundaries. One may want to argue that the lubrication theory is valid only in the limit of small tilting angles (i.e. microscopic wettabilities), and the presence of finite wettabilities can produce a small mismatch between theory and numerics. Anyhow, it should also be noted that close to the boundaries the inner physics is completely different: in the sharp interface limit the dynamics is dominated by slip properties while in the diffuse interface case by the finite thickness of the interface. Overall, at least for the range of parameters studied, the agreement is satisfactory showing a good global univarsality with respect to the mechanisms resolving the contact line singularity.

\begin{figure}
\includegraphics[scale=.32,angle=-90]{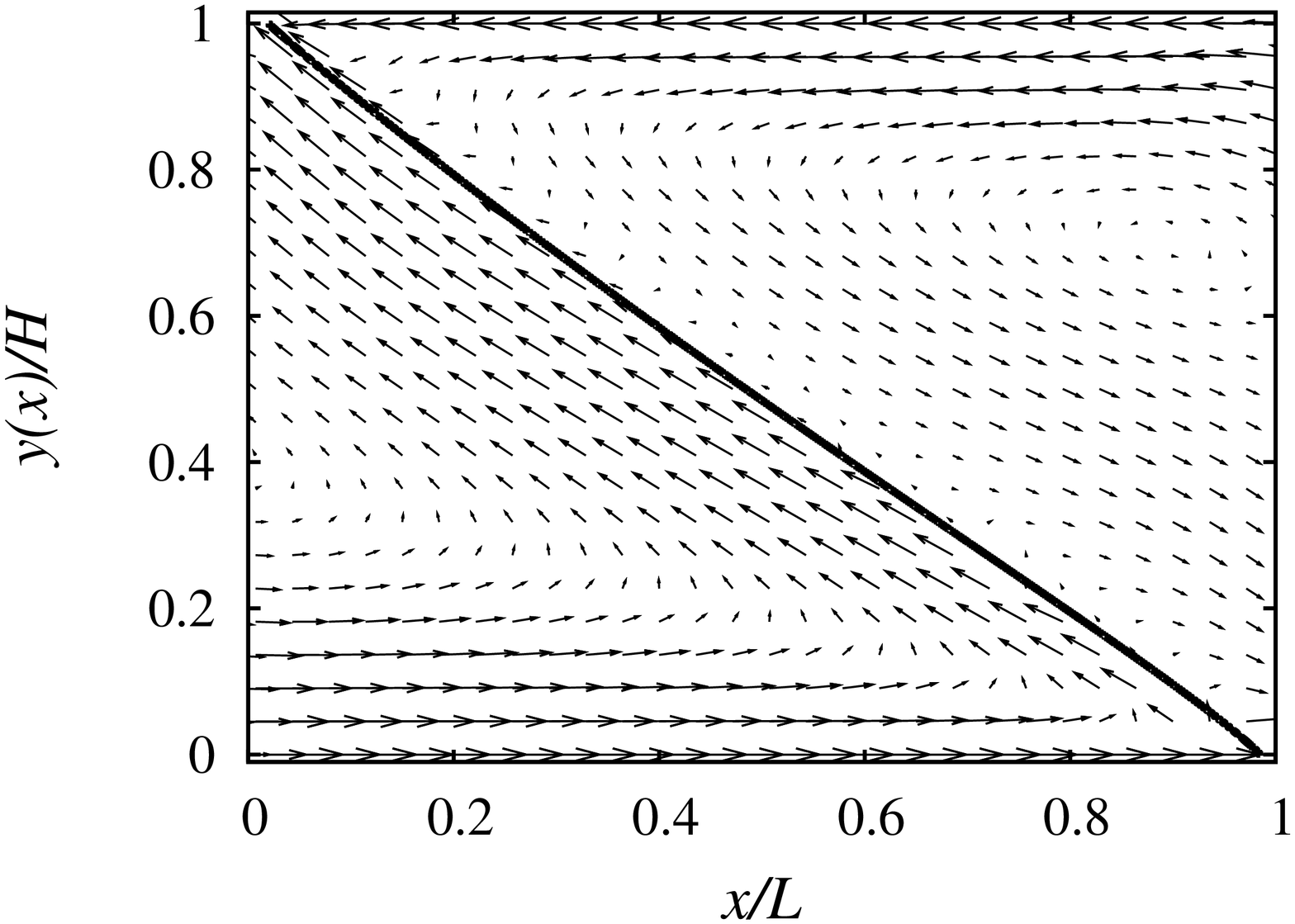}
\includegraphics[scale=.32,angle=-90]{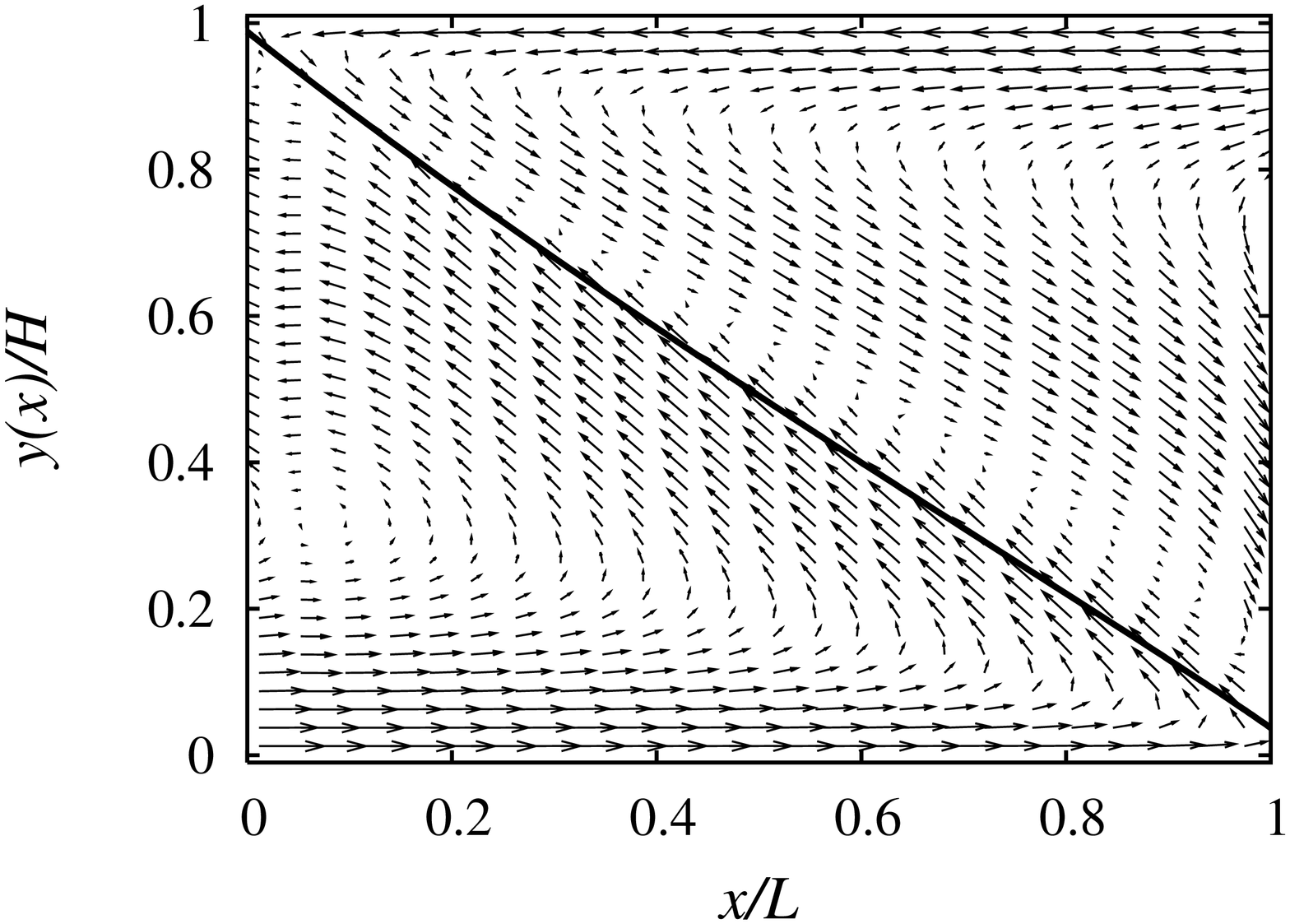}
\begin{center}
  \caption{A stationary state of the Couette cell in the numerics with diffuse interface (left) and in the sharp interface theory (right). The vertical length scale has been made dimensionless with respect to the height $H$ and the horizontal one with respect to the  horizontal extension of the interface $L$. Good agreement is found in the qualitative details of the velocity fields: a two-rolls structure is developed in both the gas and liquid phases. These plots share qualitative feature with the molecular dynamics simulations by Thompson and Robbins (1989). The interface location is also displayed (lines angling from wall to wall). }
\label{fig:12}
\end{center}
\end{figure}

\begin{figure}
\includegraphics[scale=.32,angle=-90]{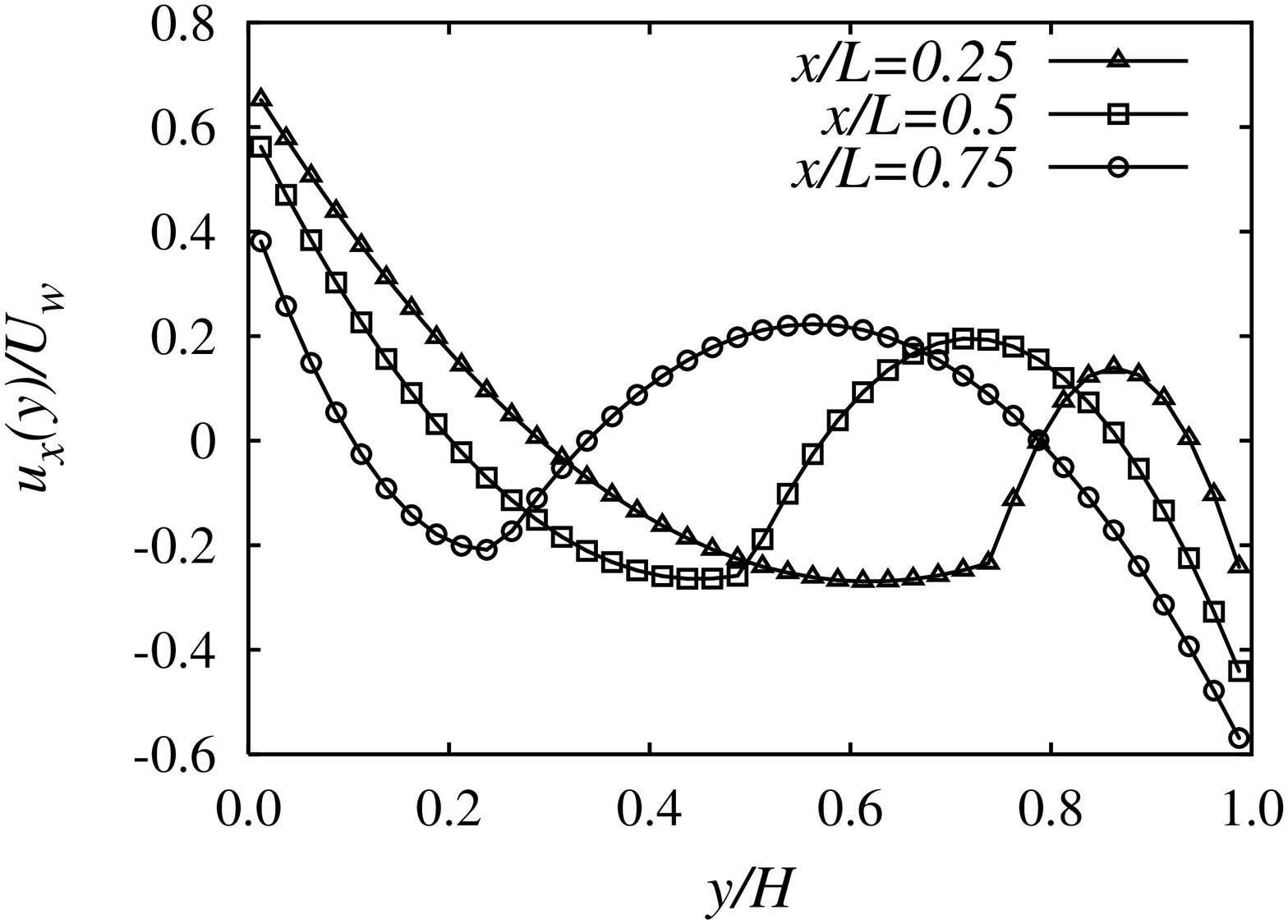}
\includegraphics[scale=.32,angle=-90]{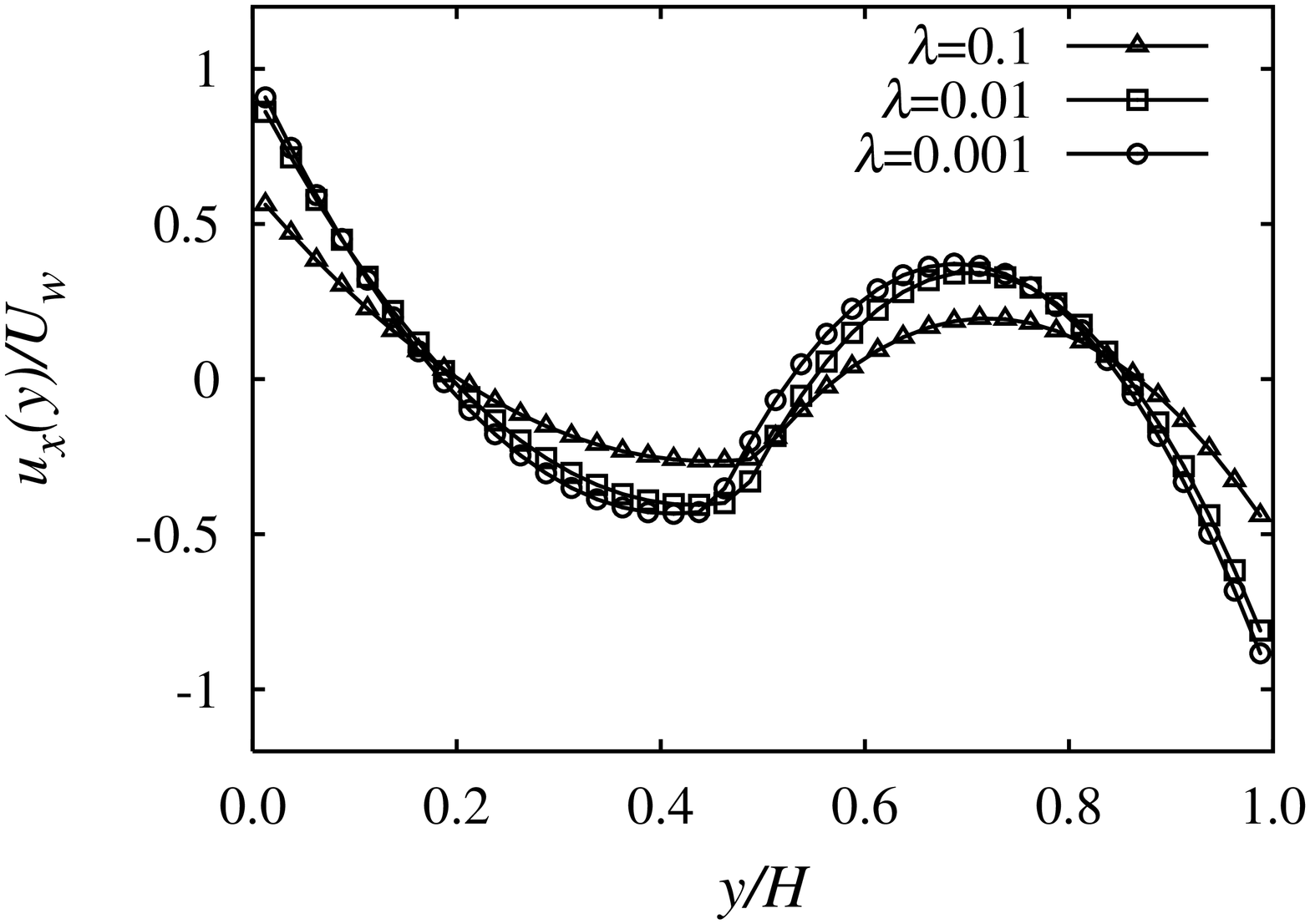}
\begin{center}
  \caption{Velocity profiles in the streamwise direction as obtained from theory. Left: velocity field $u_{x}(x,y)$ as a function of $y/H$ for different values of $x$ along the horizontal direction (see also figure \ref{fig:12}) of the interface $L$: $x/L=0.25$ (triangles), $x/L=0.5$ (squares), $x/L=0.75$ (circles). The separation of scale is $\lambda=0.1$, the capillary number is $Ca=0.011$ and the microscopic angle is $\theta_m=58^{\circ}$. Right: The velocity profiles in the streamwise direction $u_{x}(x,y)$ as a function of $y/H$, for $x/L=0.5$ and different separations of scales: $\lambda=0.1$ (triangles), $\lambda=0.01$ (squares), $\lambda=0.001$ (circles).}
\label{fig:13}
\end{center}
\end{figure}

Much more stringent comparisons can be carried out looking at the local details in the stationary regime of the velocity field. Numerical and theoretical snapshots of the velocity vector are displayed in figure \ref{fig:12}. Qualitative features are very similar for  both cases. In fact, it is evident from both theory and numerics the presence of a two-rolls structure (\cite{Thompson89}) in the 2 fluids. Looking at the details of the velocity field in the streamwise ($x$) direction and we can further compare theory and numerics. In the left panel of figure \ref{fig:13}, for a  fixed capillary number $Ca$, viscous ratio $\chi$, separation of scale $\lambda$  and  microscopic wettability $\theta_{m}$, we show the streamwise component of the velocity ($u_x (x,y)$) as a function of $y$ for various horizontal locations $x$, as extracted from the theory developed in section \ref{sec:COUETTE}   . Further details in terms of the separation of scales $\lambda$ are displayed in the right panel of the same figure. When we fix the parameters $Ca$, $\chi$ and $\theta_m$, the velocity profiles tend to converge to the same master plot for small values of $\lambda$.

\begin{figure}
\begin{center}
\includegraphics[scale=0.4,angle=-90]{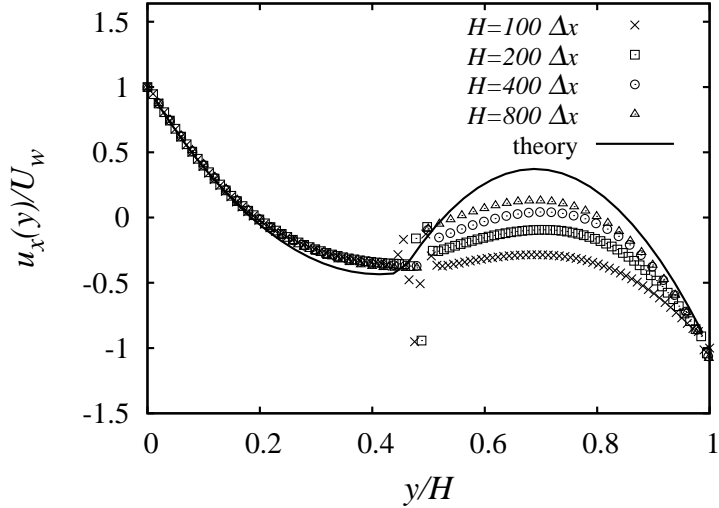}
  \caption{Comparison between theory and numerical simulations. We show the velocity field in the streamwise direction $u_{x}(x,y)$ as a function of $y/H$, for $x/L=0.5$ ($L$ is the horizontal length scale of the interface, see also figure \ref{fig:12}), a capillary number $Ca=0.011$ and a microscopic angle $\theta_{m}=58^{\circ}$. The theoretical profile is obtained with a separation of scale $\lambda=0.001$ (already in the region of saturation shown in figure \ref{fig:13}). The numerical results are obtained with a fixed interface width $\xi$ and increasing numerical resolution in the vertical direction, i.e.  decreasing the separation of scale $\lambda_{LBE}=\xi/H$. }
\label{fig:14}
\end{center}
\end{figure}

It is interesting to carry out a  similar analysis in the numerics. To do that, we can fix the viscous ratio $\chi$, the microscopic wettability $\theta_m$ and the capillary number $Ca$ to be the same as the case of the right panel of figure \ref{fig:13}. With the interface width $\xi$ fixed to a given value, we can then increase the resolution between the walls and study the way the diffuse interface velocity converges to the sharp interface prediction. Results are presented in figure \ref{fig:14}. We notice that by increasing the resolution from $H=100 \Delta x$ to $H=800 \Delta x$  we reach already for $H=600 \Delta x$ a limiting profile. On the other hand, if we compare the asymptotic numerical profile with the theoretical prediction (solid line in figure \ref{fig:14}) we observe a mismatch in the less viscous region. The discrepancy can be due to different factors. First, we can argue that the sharp interface prediction coming from the lubrication theory may not be fully correct for finite wettabilities, i.e. a small overestimate in the pressure gradients would be amplified in the less viscous part as predicted by the stationary Stokes equations (\ref{STOKES}). It clearly would be possible to clarify this point using a full 2-dimensional calculation, as for example the one discussed by \cite{Somalinga00}. Second, one should notice that in the interface region  spurious effects emerge in the numerics. This is due to the presence of a stretched interface with local gradients in the density field (\cite{Sbragaglia07,Yuan06}). Neverthless, the comparison of the shape of the interfaces between LBE and the sharp interface hydrodynamics did not reveal a great mismatch (see figures \ref{fig:9}, \ref{fig:10} and \ref{fig:11}), i.e. to this regards the effect of spurious currents is certainly small. In order to progress in this issue, one should work out a  better and refined numerical scheme to reduce the spurious currents effect and check the comparison with theory again (\cite{Sbragaglia07,Shan06,Lee06,Wagner03,Cristea03,Yuan06}).

\section{Discussions and Conclusions}

We have developed a sharp interface theory to describe a Couette cell consisting of two immiscible fluids. The two contact lines at the walls develop divergent viscous stresses and such a  singularity is removed introducing a finite slip length at the boundaries ($\ell_s$). The stationary properties of the system have been quantified in terms of the capillary number ($Ca$), the viscous ratio ($\chi$), the microscopic wettability ($\theta_m$) and the separation of scale ($\lambda$) between the inner physics ($\ell_s$) and the outer geometry (i.e. the distance between the walls $H$). The problem can be closed using the assumption of small tilting angles at the interface (lubrication approximation), thus determining the stationary interfaces in terms of $Ca$, $\chi$,  $\theta_m$ and $\lambda$. It is observed that there it exists a critical capillary number, $Ca_{cr}$, above which no stationary solution can be found. This critical capillary number corresponds to the case where liquid deposition occurs on the solid: the interface is stretched at the point that it cannot sustain any longer a stationary profile that is broken in favor of liquid entrainment on the wall. Our analysis offers the possibility to examine this critical capillary numbers in terms of all the parameters, especially the viscous ratio, thus extending a similar analysis done in a  recent paper by \cite{Jacqmin04}.  Moreover, in the limit  of  small viscous ratios, it allows to compare our results with the recent analysis proposed by \cite{Eggers04} for the case of a plate withdrawn at a given speed from a liquid bath. This elucidates the role of completely different geometries in determining the critical capillary numbers.\\
In the second part of the paper we have studied the same system with a diffuse interface model based on the Lattice Boltzmann equation (LBE). The comparison with LBE results allows us to benchmark the numerical model and to understand the effects of finite thickness of the interface on global quantities, as the critical capillary number, as well as local ones, as the interface shape and the velocity profiles. Good agreement is found when the scale separation in the LBE, given by the ratio between interface thickness $\xi$ and distance between the walls $H$, becomes small enough.\\ 
Extensions of the LBE simulations to other  geometries, as it is the case for the Landau-Levich case in presence of gravity, would allow for direct comparison with experimental results (\cite{Quere91,Snoeijer06,Snoeijer07}) and also offer some more physical insight in problems that are still not completely understood, as the importance of roughness, contact angle hysteresis and speed dependency of the microscopic wetting properties (\cite{Quere91,Golestanian03,Rame04,Heine04}). \\
Very recently, a withdrawal experiment by Snoeijer {\it et al.} (\cite{Snoeijer06}) has been performed and, contrary to previous experiments (\cite{Sedev91}), the transition to film entrainment did not occur at the critical capillary number predicted by the theory (\cite{Eggers04,Eggers05}). This avoided critical bahaviour is due to the formation of a  capillary ridge that does not trivially match the liquid film and  determine the critical speed of entrainment. This can be related to contact angle hysteresis that has not been treated in the continuum models: the sensitivity of $Ca_{cr}$ with respect to $\theta_m$ (see also our figure \ref{fig:3}) can also support this interpretation. Another interesting possibility would be the introduction of a mesoscopic roughness in the numerical simulations and study the way this critical behaviour changes in terms of the underlying heterogeneities (\cite{Kusumaatmaja07}). One has to caution that such a kind of analysis would be limited to finite viscous ratios, unless one is opting for more sophisticated schemes incorporating density ratios comparable with liquid-gas interfaces used in the experiments (\cite{Inamuro04,Lee05}). Moreover, the properties of the diffuse interface models will emerge as a  function of the separation of scale $\lambda_{LBE}$:  reaching  extremely small values of $\lambda_{LBE}$ is somehow prohibitive because of the extremely large resolution needed to simulate a set of scales ranging from $nm$ to tens of  $\mu m$. To this regard, similar analysis to that presented in this paper (see section \ref{sec:visco}) would help to translate the numerical observations into realistic numbers.

\begin{acknowledgments}
We are indebted with B. Andreotti, R. Benzi, J. Yeomans, H. Kusumaatmaja, F. Toschi and S. Succi for useful and enlightening discussions. 
\end{acknowledgments}

\appendix

\section{Matrix problem for the lubrication approximation}

In this appendix we describe in details the calculations needed to solve the Couette problem in the lubrication approximation. Consistently with the assumption of small tilting angles, the normal vector to the interface is 
\be
\hat{n} = \frac{1}{\sqrt{1+\left( \frac{d h}{d x} \right)^2}} \left( \hat{e}_{y}-\hat{e}_{x} \frac{d h}{d x} \right)
\ee
with $\hat{e}_{x}$ and $\hat{e}_{y}$ Cartesian unit vectors. A zero normal component for the left field (for the right one similar arguments can be applied)  at the interface means
\be\label{NORM}
\left. u_{l,n} \right|_{h}=\left. u_{y,l}\right|_{h}-\frac{d h}{d x} \left. u_{x,l} \right|_{h}=0.
\ee
Moreover from the continuity equation (\ref{CCC}) we derive that
\be\label{V2}
\left. u_{y,l} \right|_{h}=-\frac{d}{dx }\int_{0}^{h(x)} u_{x,l} dy+\frac{d h}{d x} \left. u_{x,l} \right|_{h}. 
\ee
The use of equation (\ref{V2}) together with equation (\ref{NORM}) leads to
\be
\left. u_{l,n} \right|_{h}=\frac{d}{dx }\int_{0}^{h(x)} u_{x,l} dy. 
\ee
Since in the  inner (outer) region  of the system there is no net mass flow rate, this means that
\be\label{NOFLUX}
\int_{0}^{h(x)} u_{x,l} dy=0\hspace{.2in}
\int_{h(x)}^{H} u_{x,r} dy=0.
\ee
The 6 boundary conditions that we need to fix $A_{l}$,$B_{l}$,$p_{l,x}$,$A_{r}$,$B_{r}$,$p_{l,x}$ are then represented by (\ref{CONT}),(\ref{SLIP}) and (\ref{NOFLUX}). They  translate in the following system:
\be\label{SYS:1}
A_{l}+h B_{l}+\frac{h^2}{2} p_{l,x}-\chi^{-1} \left( A_{r}+h B_{r}+\frac{h^2}{2}p_{r,x}\right)=0
\ee
\be\label{SYS:2}
B_{l}+h p_{l,x}- B_{r}-h p_{r,x} =0
\ee
\be\label{SYS:3}
A_{l}-\ell_{s}B_{l}=\mu_{l} U_{w}
\ee
\be\label{SYS:4}
A_{r}+(H+\ell_{s})B_{r}+\left( \frac{H^2}{2}+\ell_{s}H \right)p_{r,x}=-\mu_{r} U_{w}
\ee
\be\label{SYS:5}
h A_{l}+\frac{h^2}{2}B_{l}+\frac{h^{3}}{6}p_{l,x}=0
\ee
\be\label{SYS:6}
(H-h)A_{r}+\frac{1}{2}(H^2-h^2)B_{r}+\frac{1}{6}(H^3-h^3)p_{r,x}=0.
\ee

\end{document}